\documentclass[11pt,a4paper]{article}
\usepackage[T1]{fontenc}
\usepackage{lmodern}
\input{glyphtounicode}
\usepackage[margin=2.5cm]{geometry}
\usepackage{amsmath,amssymb,amsthm}
\usepackage{tabularray}
\usepackage{tikz}
\tikzset{
    dtroot/.style = {circle, fill, inner sep = 1.6pt},
    dtquery/.style = {circle, draw, line width = 0.4pt, inner sep = 0.5pt, minimum size = 8.5mm},
    dtleaf/.style = {rectangle, draw, line width = 0.4pt, rounded corners = 1pt, inner sep = 3pt, minimum height = 6mm},
    dtthin/.style = {line width = 0.4pt},
    dtlab/.style = {font = \normalsize, inner sep = 2pt},
    dtpath/.style = {line width = 1.4pt}
}
\usepackage[hidelinks]{hyperref}

\newtheorem{theorem}{Theorem}
\newtheorem{lemma}{Lemma}
\newtheorem{proposition}{Proposition}
\newtheorem{corollary}{Corollary}

\theoremstyle{definition}
\newtheorem{definition}{Definition}
\newtheorem{example}{Example}
\theoremstyle{remark}
\newtheorem{remark}{Remark}

\newcommand{\Str}[1]{\ensuremath{\text{\textrm{\texttt{#1}}}}}

\newcommand{\Stra}{\Str{a}}
\newcommand{\Strb}{\Str{b}}
\newcommand{\Strc}{\Str{c}}
\newcommand{\Strzero}{\Str{0}}
\newcommand{\Strone}{\Str{1}}
\newcommand{\Strtwo}{\Str{2}}

\DeclareMathOperator{\Hom}{Hom}
\DeclareMathOperator{\Het}{Het}
\DeclareMathOperator{\Av}{Av}

\newcommand{\hrd}{h^{\mathrm{rd}}}
\newcommand{\hra}{h^{\mathrm{ra}}}
\newcommand{\hmd}{h^{\mathrm{md}}}
\newcommand{\hma}{h^{\mathrm{ma}}}
\newcommand{\Lsort}{L_{\mathrm{sort}}}
\newcommand{\Shom}{S_{\mathrm{hom}}}
\newcommand{\Shet}{S_{\mathrm{het}}}

\begin{document}

\title{Decision trees for subword-closed languages over arbitrary finite alphabets}
\author{Shahid Hussain\\[2pt] \small Department of Computer Science, Institute of Business Administration (IBA), Karachi, Pakistan\\ \small \texttt{shahidhussain@iba.edu.pk}}
\date{20 November 2026}
\maketitle

\begin{abstract}
Moshkov classified binary subword-closed languages $L$ by the minimum depth of decision trees that recognize the words of length $n$ in $L$, or decide membership in $L$, using queries that return the $i$th letter of the input word. We extend this classification to alphabets with $k \geq 2$ letters. With the homogeneity and heterogeneity dimensions defined via pairs of different letters, the classification is the same as for $k = 2$: the deterministic recognition depth is bounded, logarithmic, or linear in $n$, the other three depths are bounded or linear, and there are five complexity classes. Since Moshkov's two-block structure lemma fails for $k \geq 3$, we prove a structure lemma with up to $k$ blocks. Our bounds are polynomial in $k$ and the two dimensions; in the bounded case they replace the exponential constant of the binary case by a linear one. We also show that the complexity class (one of the five) of a language given by a finite set of forbidden subwords can be computed in polynomial time. For several ternary languages we prove exact formulas for all four depths and check them by computation for small $n$.
\end{abstract}

\section{Introduction}

A language $L$ over a finite alphabet $\Sigma$ is \emph{subword closed} if the word $u_{1} \cdots u_{m}$ belongs to $L$ whenever a word of the form $w_{1}u_{1}w_{2} \cdots w_{m}u_{m}w_{m+1}$ belongs to $L$. Subword-closed languages, also called downward-closed languages with respect to the scattered subword order, are studied in formal language theory from several points of view; see, for example, \cite{AtminasLozin2024,BrzozowskiJiraskovaZou2014,Haines1969,Hospodar2021,Okhotin2010}. By the results of Higman~\cite{Higman1952} and Haines~\cite{Haines1969}, every subword-closed language over a finite alphabet is regular; more precisely, it is the set of words avoiding (as subwords) the words of a finite set.

Moshkov introduced the minimum depth of deterministic and nondeterministic decision trees for the recognition and membership problems as new parameters of formal languages \cite{Moshkov1997,Moshkov2000,Moshkov2005,Moshkov2022}. For a language~$L$ and a natural number $n$, let $L(n)$ be the set of words of length $n$ in~$L$. In the \emph{recognition problem}, for a given word of $L(n)$, we must determine this word; in the \emph{membership problem}, for a given word of length $n$ over $\Sigma$, we must decide whether it belongs to $L(n)$. In both problems we may ask queries, each of which returns the $i$th letter of the input word for some $i \in \{1, \ldots, n\}$. For regular languages, the behavior of the minimum depth of such trees as $n$ grows was described in \cite{Moshkov1997,Moshkov2000,Moshkov2005}; these results, however, require a deterministic finite automaton accepting the language. General methods for bounding the depth of deterministic and nondeterministic decision trees are developed in \cite{Moshkov2005,Moshkov2020}; for the related notions of query and certificate complexity of Boolean functions see \cite{BuhrmanDeWolf2002}.

In \cite{Moshkov2023}, Moshkov considered binary subword-closed languages, i.e., the case $\Sigma = \{\Strzero, \Strone\}$, without assuming that an automaton is given. He introduced two combinatorial parameters, the homogeneity dimension $\Hom(L)$ and the heterogeneity dimension $\Het(L)$, and proved that they determine the asymptotic behavior of the minimum depth of decision trees: for deterministic recognition the depth is $O(1)$, $\Theta(\log n)$, or $\Theta(n)$, and for nondeterministic recognition and for deterministic and nondeterministic membership it is $O(1)$ or $\Theta(n)$. The joint behavior of the four depths yields five complexity classes of binary subword-closed languages. These results are also presented in the book \cite[pp.~213--222]{DurdymyradovMoshkovOstonov2024}.

In this paper we generalize these results to an arbitrary finite alphabet with $k \geq 2$ letters.\footnote{The unary case $k = 1$ is trivial: a subword-closed language over $\{\Stra\}$ is $\emptyset$, $\{\Stra^{i} : i \leq M\}$ for some $M$, or $\Stra^{*}$, and all four depths are $0$ for every $n$, since $\lvert \Sigma^{n} \rvert = 1$.} The classification and the five complexity classes are the same as in the binary case, and the size of the alphabet affects only the constants. However, the binary proof does not extend to $k \geq 3$, and new arguments are needed.

The novelty of the results considered in this paper is connected with the following. First, we define $\Hom(L)$ and $\Het(L)$ for arbitrary alphabets (Definitions~\ref{def:hom} and~\ref{def:het}); for $k = 2$ these definitions coincide with Moshkov's. Second, we show that the structure lemma \cite[Lemma~1]{Moshkov2023}, which represents every word of $L$ with at most two long blocks of letters, fails for $k \geq 3$, and we replace it by a structure lemma with at most $k$ long blocks (Lemma~\ref{lem:structure}); the number $k$ is attained (Proposition~\ref{prop:sorted}). Third, we prove that Moshkov's classification holds for every $k \geq 2$ (Theorems~\ref{thm:recognition} and~\ref{thm:membership}), with explicit bounds that are polynomial in $k$ and $\Hom(L)$, and in the bounded case also in $\Het(L)$. In the bounded case the depth of deterministic recognition is at most $2k(2\Hom(L) + 1) + 2(k - 1)\Het(L) + 2$, while the binary argument gives a constant that is exponential in $\Hom(L)$ and $\Het(L)$. The logarithmic upper bound is obtained by a binary search over segments of the word (Lemma~\ref{lem:logdepth}); if $\Hom(L) = 0$, the same procedure gives the bound $2 + (k - 1)\lceil \log_{2} n \rceil$ (Proposition~\ref{prop:hom0}). Fourth, for the language of sorted words over $k$ letters, the deterministic recognition depth lies between $(k - 1)\log_{k}(n/(k - 1))$ and $(k - 1)\lceil \log_{2}(n + 1) \rceil$ (Proposition~\ref{prop:sorted}), so the constant in the logarithmic case grows with $k$. Fifth, we describe the five complexity classes for every $k$ (Table~\ref{tab:classes} and Proposition~\ref{prop:classes}) and show that the class of a language given by a finite set of forbidden subwords can be found in polynomial time (Proposition~\ref{prop:effective}). Finally, for six ternary languages we find all four depths exactly for every $n$ (Proposition~\ref{prop:exact}). For the language $\{\Strzero^{i}\Strone^{j}\Strtwo^{l}\}$, the deterministic recognition depth satisfies an exact recursion and is equal to $\lfloor \log_{2} n \rfloor + \lfloor \log_{2}(n/3) \rfloor + 2$ for $n \geq 2$ (Proposition~\ref{prop:recursion}). These values were also confirmed by exhaustive computation.

This paper consists of six sections. Section~\ref{sec:notions} contains the main notions. Section~\ref{sec:results} contains the main results and examples, and Section~\ref{sec:proofs} contains their proofs. In Section~\ref{sec:comparison} we compare our results and constants with the binary case. Section~\ref{sec:conclusions} contains conclusions and open problems.

\section{Main notions}\label{sec:notions}

Let $\mathbb{N} = \{0, 1, 2, \ldots\}$ and let $\Sigma$ be a finite alphabet with $\lvert \Sigma \rvert = k \geq 2$. By $\Sigma^{*}$ we denote the set of all finite words over $\Sigma$, including the empty word $\varepsilon$. For a word $w \in \Sigma^{*}$, $\lvert w \rvert$ is its length and $\lvert w \rvert_{\Stra}$ is the number of occurrences of the letter $\Stra \in \Sigma$ in $w$. For $\Stra \in \Sigma$ and $i \in \mathbb{N}$, $\Stra^{i}$ is the word consisting of $i$ copies of $\Stra$, with $\Stra^{0} = \varepsilon$. A word $u$ is a \emph{subword} of a word $w$ if $u$ can be obtained from $w$ by deleting some (possibly none) of its letters, and $u$ is a \emph{factor} of $w$ if $w = xuy$ for some words $x, y$; thus factors are contiguous, while subwords need not be. A subset $L \subseteq \Sigma^{*}$ is called a \emph{language} over $\Sigma$, and $L^{C} = \Sigma^{*} \setminus L$ is its complementary language. The notation $\lvert L \rvert = \infty$ means that $L$ is infinite, and $\lvert L \rvert < \infty$ means that $L$ is finite. For a set $X$ of words, $\mathord{\downarrow}X$ denotes the set of all subwords of words of $X$.

\begin{definition}[Subword-closed language]
A language $L \subseteq \Sigma^{*}$ is \emph{subword closed} if, for every $m \geq 1$ and all words $w_{1}, \ldots, w_{m+1}, u_{1}, \ldots, u_{m} \in \Sigma^{*}$, the condition $w_{1} u_{1} w_{2} \cdots w_{m} u_{m} w_{m+1} \in L$ implies $u_{1} \cdots u_{m} \in L$.
\end{definition}

Equivalently, every subword of a word of $L$ belongs to $L$: a word $u$ is a subword of $w$ if and only if $w = w_{1}u_{1}w_{2} \cdots w_{m}u_{m}w_{m+1}$ with $u = u_{1} \cdots u_{m}$ for some words $w_{i}, u_{j}$ (take the $u_{j}$ to be single letters, or $m = 1$ and $u_{1} = \varepsilon$ for $u = \varepsilon$).

\begin{definition}[Avoidance language]\label{def:av}
For a set $F \subseteq \Sigma^{*}$ of \emph{forbidden words}, the \emph{avoidance language} of $F$ is
\[
    \Av(F) \;=\; \bigl\{\, w \in \Sigma^{*} \;\bigm|\; \text{no word of } F \text{ is a subword of } w \,\bigr\}.
\]
\end{definition}

The language $\Av(F)$ is subword closed for every $F$: if $u$ is a subword of $w \in \Av(F)$ and some $f \in F$ were a subword of $u$, then $f$ would be a subword of $w$, since the subword relation is transitive. Conversely, every subword-closed language $L$ satisfies $L = \Av(L^{C})$, and also $L = \Av(F)$, where $F$ is the set of minimal words of $L^{C}$ with respect to the subword order. By Higman's lemma~\cite{Higman1952}, every set of pairwise incomparable words is finite, so $F$ is finite. In particular, $\Av(\emptyset) = \Sigma^{*}$, and $\Av(F) = \emptyset$ if $\varepsilon \in F$.

Throughout, $n$ denotes a natural number with $n \geq 1$. Let $\Sigma^{n}$ be the set of words of length~$n$ over $\Sigma$ and $L(n) = L \cap \Sigma^{n}$. For $i \in \{1, \ldots, n\}$, the \emph{attribute} (query) $l^{n}_{i} \colon \Sigma^{n} \to \Sigma$ is defined by $l^{n}_{i}(a_{1} \cdots a_{n}) = a_{i}$; that is, $l^{n}_{i}$ returns the $i$th letter of the word. Put $A_{n} = \{l^{n}_{1}, \ldots, l^{n}_{n}\}$. We consider two problems related to $L(n)$.
\begin{itemize}
    \item \emph{Recognition}: for a given word $w \in L(n)$, determine $w$ using attributes from $A_{n}$.
    \item \emph{Membership}: for a given word $w \in \Sigma^{n}$, determine whether $w \in L(n)$ using attributes from~$A_{n}$.
\end{itemize}
Recognition is a promise problem: the input is known to belong to $L(n)$. Membership is defined for every input in $\Sigma^{n}$.

\begin{definition}[Decision tree]
A \emph{decision tree over $L(n)$} is a finite directed tree with a root satisfying the following conditions:
\begin{enumerate}
    \item the root and the edges leaving the root are not labeled;
    \item each node that is neither the root nor a terminal node is labeled with an attribute from~$A_{n}$;
    \item each edge leaving a node other than the root is labeled with a letter from $\Sigma$.
\end{enumerate}
The labels of terminal nodes depend on the problem being solved and are specified in Definitions~\ref{def:recognition} and~\ref{def:membership}.
\end{definition}

The root plays the role of a starting point and asks no query. In a nondeterministic tree, the several edges leaving the root correspond to different ``guesses''; in a deterministic tree, the unique edge leaving the root leads to the node with the first query (or directly to a terminal node). A nondeterministic tree does not accept or reject inputs: by Definitions~\ref{def:recognition} and~\ref{def:membership} below, every complete path that is compatible with an admissible input has to end in the correct answer for this input.

\begin{definition}[Deterministic decision tree]
A decision tree over $L(n)$ is \emph{deterministic} if exactly one edge leaves the root and, for each node that is neither the root nor a terminal node, the edges leaving this node are labeled with pairwise different letters. In particular, every such node has at most $k$ outgoing edges. Decision trees in general (without these conditions) are also called \emph{nondeterministic}.
\end{definition}

\begin{definition}[Complete path]
Let $\Gamma$ be a decision tree over $L(n)$. A \emph{complete path} in $\Gamma$ is a sequence $\xi = v_{0}, e_{0}, v_{1}, e_{1}, \ldots, v_{m}, e_{m}, v_{m+1}$ of nodes and edges of $\Gamma$ such that $v_{0}$ is the root, $v_{m+1}$ is a terminal node, and, for $j = 0, \ldots, m$, the edge $e_{j}$ leaves $v_{j}$ and enters $v_{j+1}$. If $m = 0$, put $\Sigma^{n}(\xi) = \Sigma^{n}$. If $m > 0$, let the node $v_{j}$ be labeled with the attribute $l^{n}_{i_{j}}$ and the edge $e_{j}$ be labeled with the letter $\Strb_{j}$ for $j = 1, \ldots, m$, and put
\[
    \Sigma^{n}(\xi) \;=\; \bigl\{\, a_{1} \cdots a_{n} \in \Sigma^{n} \;\bigm|\; a_{i_{1}} = \Strb_{1}, \ldots, a_{i_{m}} = \Strb_{m} \,\bigr\}.
\]
\end{definition}

\begin{definition}[Recognition problem]\label{def:recognition}
Let $L(n)$ be nonempty. A decision tree $\Gamma$ over $L(n)$ \emph{solves the problem of recognition for $L(n)$ nondeterministically} if
\begin{enumerate}
    \item each terminal node of $\Gamma$ is labeled with a word from $L(n)$;
    \item for each $w \in L(n)$, there exists a complete path $\xi$ in $\Gamma$ such that $w \in \Sigma^{n}(\xi)$;
    \item for each $w \in L(n)$ and each complete path $\xi$ in $\Gamma$ such that $w \in \Sigma^{n}(\xi)$, the terminal node of $\xi$ is labeled with $w$.
\end{enumerate}
The tree $\Gamma$ \emph{solves the problem of recognition for $L(n)$ deterministically} if $\Gamma$ is deterministic and solves the problem of recognition for $L(n)$ nondeterministically.
\end{definition}

\begin{definition}[Membership problem]\label{def:membership}
A decision tree $\Gamma$ over $L(n)$ \emph{solves the problem of membership for $L(n)$ nondeterministically} if
\begin{enumerate}
    \item each terminal node of $\Gamma$ is labeled with a number from $\{0, 1\}$;
    \item for each $w \in \Sigma^{n}$, there exists a complete path $\xi$ in $\Gamma$ such that $w \in \Sigma^{n}(\xi)$;
    \item for each $w \in \Sigma^{n}$ and each complete path $\xi$ in $\Gamma$ such that $w \in \Sigma^{n}(\xi)$, the terminal node of $\xi$ is labeled with $1$ if $w \in L(n)$ and with $0$ otherwise.
\end{enumerate}
The tree $\Gamma$ \emph{solves the problem of membership for $L(n)$ deterministically} if $\Gamma$ is deterministic and solves the problem of membership for $L(n)$ nondeterministically.
\end{definition}

The terminal labels $0$ and $1$ in Definition~\ref{def:membership} are answers (\emph{no} and \emph{yes}) and should not be confused with letters of $\Sigma$, even when $\Strzero, \Strone \in \Sigma$.

\begin{definition}[Depth]\label{def:depth}
The \emph{depth} $h(\Gamma)$ of a decision tree $\Gamma$ over $L(n)$ is the maximum number of nodes in a complete path of $\Gamma$ that are neither the root nor the terminal node. We denote by $\hra_{L}(n)$ (respectively, $\hrd_{L}(n)$) the minimum depth of a decision tree over $L(n)$ that solves the problem of recognition for $L(n)$ nondeterministically (respectively, deterministically), and by $\hma_{L}(n)$ (respectively, $\hmd_{L}(n)$) the minimum depth of a decision tree over $L(n)$ that solves the problem of membership for $L(n)$ nondeterministically (respectively, deterministically). If $L(n) = \emptyset$, then $\hra_{L}(n) = \hrd_{L}(n) = \hma_{L}(n) = \hmd_{L}(n) = 0$.
\end{definition}

In the notation of \cite{Moshkov2023}, the first letter of each superscript refers to the problem (r for recognition, m for membership) and the second to the type of tree (d for deterministic, a for arbitrary, i.e., not necessarily deterministic). For $\Sigma = \{\Strzero, \Strone\}$, the notions above coincide with those of \cite{Moshkov2023}. Since every deterministic decision tree is a decision tree, $\hra_{L}(n) \leq \hrd_{L}(n)$ and $\hma_{L}(n) \leq \hmd_{L}(n)$. Moreover, a deterministic decision tree $\Gamma$ has at most $k^{h(\Gamma)}$ terminal nodes, because exactly one edge leaves the root and at most $k$ edges leave every other nonterminal node. Finally, a decision tree computing $l^{n}_{1}, \ldots, l^{n}_{n}$ one after another shows that all four depths are at most $n$.

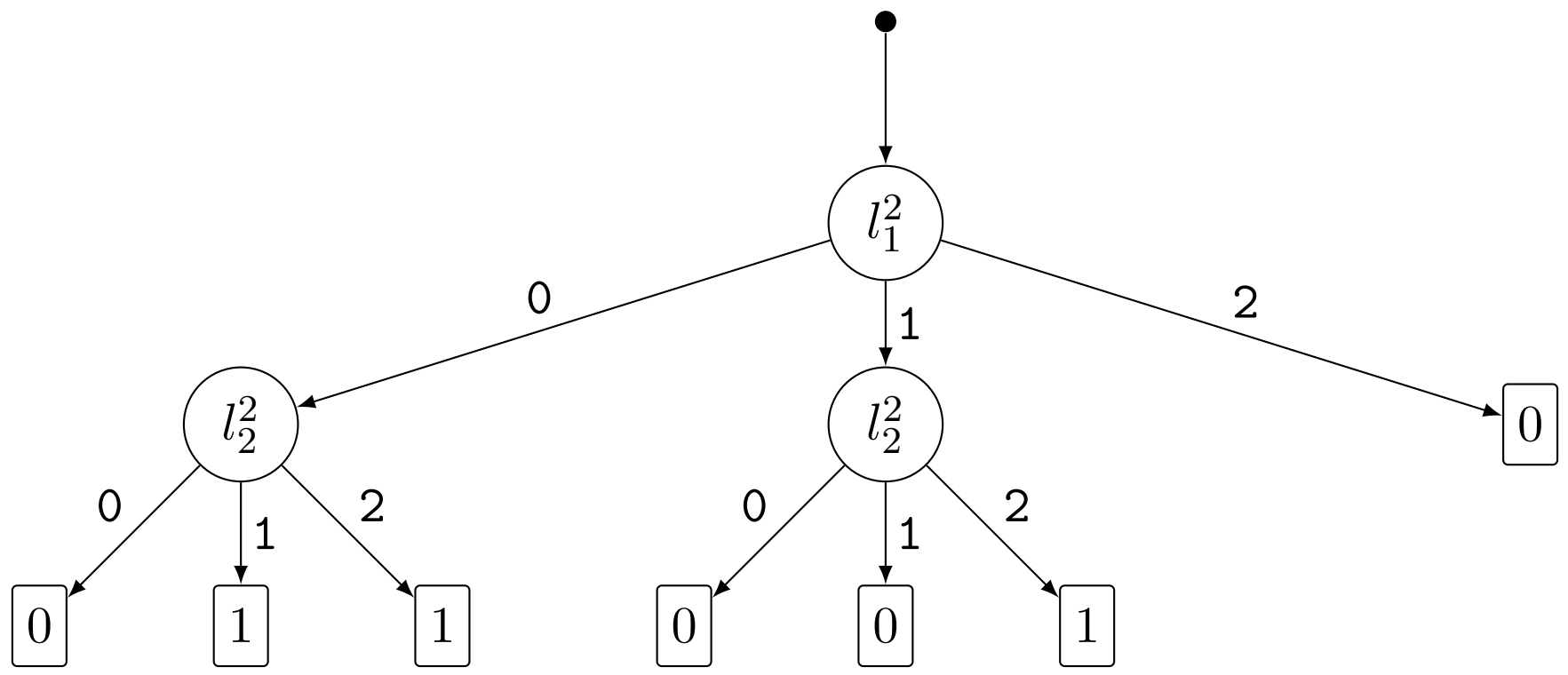
\begin{figure}[ht]
\centering
\begin{tikzpicture}[level distance = 15mm, edge from parent/.style = {draw, -latex}]
\node[dtroot] {}
    child {node[dtquery] {$l^{2}_{1}$}
        [sibling distance = 48mm]
        child {node[dtquery] {$l^{2}_{2}$}
            [sibling distance = 15mm]
            child {node[dtleaf] {$0$} edge from parent node[dtlab, above left] {$\Strzero$}}
            child {node[dtleaf] {$1$} edge from parent node[dtlab, right] {$\Strone$}}
            child {node[dtleaf] {$1$} edge from parent node[dtlab, above right] {$\Strtwo$}}
            edge from parent node[dtlab, above left] {$\Strzero$}}
        child {node[dtquery] {$l^{2}_{2}$}
            [sibling distance = 15mm]
            child {node[dtleaf] {$0$} edge from parent node[dtlab, above left] {$\Strzero$}}
            child {node[dtleaf] {$0$} edge from parent node[dtlab, right] {$\Strone$}}
            child {node[dtleaf] {$1$} edge from parent node[dtlab, above right] {$\Strtwo$}}
            edge from parent node[dtlab, right] {$\Strone$}}
        child {node[dtleaf] {$0$} edge from parent node[dtlab, above right] {$\Strtwo$}}};
\end{tikzpicture}
\caption{A deterministic decision tree of depth $2$ that solves the problem of membership for $L(2) = \{\Strzero\Strone, \Strzero\Strtwo, \Strone\Strtwo\}$, where $L = \mathord{\downarrow}\{\Strzero\Strone\Strtwo\}$ (Example~\ref{ex:trees}). Edges are labeled with letters, and terminal nodes with the answers $1$ (the word belongs to $L(2)$) and $0$ (it does not). If the first letter is $\Strtwo$, the answer is $0$ without a second query.}
\label{fig:membership}
\end{figure}

\begin{example}\label{ex:trees}
Let $\Sigma = \{\Strzero, \Strone, \Strtwo\}$, let $n \geq 2$, and let $L$ be a language with
\[
    L(n) \;=\; \{\Strzero^{n},\ \Strzero^{n-1}\Strone,\ \Strzero^{n-1}\Strtwo,\ \Strzero^{n-2}\Strone\Strtwo\};
\]
this is the language $L_{4}$ of Table~\ref{tab:examples} below. Figure~\ref{fig:recognition} shows two decision trees that solve the problem of recognition for $L(n)$. In both trees the root is drawn as a small black dot, the nodes labeled with attributes as circles, and the terminal nodes as rectangles. The tree in Figure~\ref{fig:recognition}(a) is deterministic and has depth $2$; the highlighted complete path $\xi$ has $\Sigma^{n}(\xi) = \{a_{1} \cdots a_{n} \in \Sigma^{n} : a_{n-1} = \Strzero,\ a_{n} = \Strtwo\}$, and $\Strzero^{n-1}\Strtwo$ is the only word of $L(n)$ in this set. The tree in Figure~\ref{fig:recognition}(b) is nondeterministic: four edges leave the root, and the path starting with the $j$th edge queries a recognition certificate of the $j$th word of $L(n)$ in the order listed above (see Proposition~\ref{prop:certificates}). Its depth is also $2$, but the words $\Strzero^{n}$ and $\Strzero^{n-1}\Strone$ are recognized after a single query. Figure~\ref{fig:membership} shows a deterministic decision tree of depth $2$ that solves the problem of membership for $L(2)$, where $L = \mathord{\downarrow}\{\Strzero\Strone\Strtwo\}$ and $L(2) = \{\Strzero\Strone, \Strzero\Strtwo, \Strone\Strtwo\}$.
\end{example}

\begin{figure}[ht]
\centering
\begin{minipage}[b]{0.47\textwidth}
\centering
\begin{tikzpicture}[level distance = 15mm, edge from parent/.style = {draw, -latex}]
\node[dtroot] {}
    child {node[dtquery] {$l^{n}_{n-1}$}
        [sibling distance = 32mm]
        child {node[dtquery] {$l^{n}_{n}$}
            [sibling distance = 18mm]
            child {node[dtleaf] {$\Strzero^{n}$} edge from parent[dtthin] node[dtlab, above left] {$\Strzero$}}
            child {node[dtleaf] {$\Strzero^{n-1}\Strone$} edge from parent[dtthin] node[dtlab, right] {$\Strone$}}
            child {node[dtleaf] {$\Strzero^{n-1}\Strtwo$} edge from parent[dtpath] node[dtlab, above right] {$\Strtwo$}}
            edge from parent[dtpath] node[dtlab, above left] {$\Strzero$}}
        child {node[dtleaf] {$\Strzero^{n-2}\Strone\Strtwo$} edge from parent[dtthin] node[dtlab, above right] {$\Strone$}}
        edge from parent[dtpath]};
\end{tikzpicture}

(a) Deterministic
\end{minipage}
\hfill
\begin{minipage}[b]{0.50\textwidth}
\centering
\begin{tikzpicture}[level distance = 15mm, sibling distance = 18mm, edge from parent/.style = {draw, -latex}]
\node[dtroot] {}
    child {node[dtquery] {$l^{n}_{n}$}
        child {node[dtleaf] {$\Strzero^{n}$} edge from parent node[dtlab, right] {$\Strzero$}}}
    child {node[dtquery] {$l^{n}_{n}$}
        child {node[dtleaf] {$\Strzero^{n-1}\Strone$} edge from parent node[dtlab, right] {$\Strone$}}}
    child {node[dtquery] {$l^{n}_{n-1}$}
        child {node[dtquery] {$l^{n}_{n}$}
            child {node[dtleaf] {$\Strzero^{n-1}\Strtwo$} edge from parent node[dtlab, right] {$\Strtwo$}}
            edge from parent node[dtlab, right] {$\Strzero$}}}
    child {node[dtquery] {$l^{n}_{n-1}$}
        child {node[dtleaf] {$\Strzero^{n-2}\Strone\Strtwo$} edge from parent node[dtlab, right] {$\Strone$}}};
\end{tikzpicture}

(b) Nondeterministic
\end{minipage}
\caption{Two decision trees of depth $2$ that solve the problem of recognition for $L(n) = \{\Strzero^{n}, \Strzero^{n-1}\Strone, \Strzero^{n-1}\Strtwo, \Strzero^{n-2}\Strone\Strtwo\}$, $n \geq 2$ (Example~\ref{ex:trees}). Each edge below a query is labeled with the answer, i.e., the letter at the queried position. The thick path in (a) is the complete path $\xi$ discussed in Example~\ref{ex:trees}.}
\label{fig:recognition}
\end{figure}
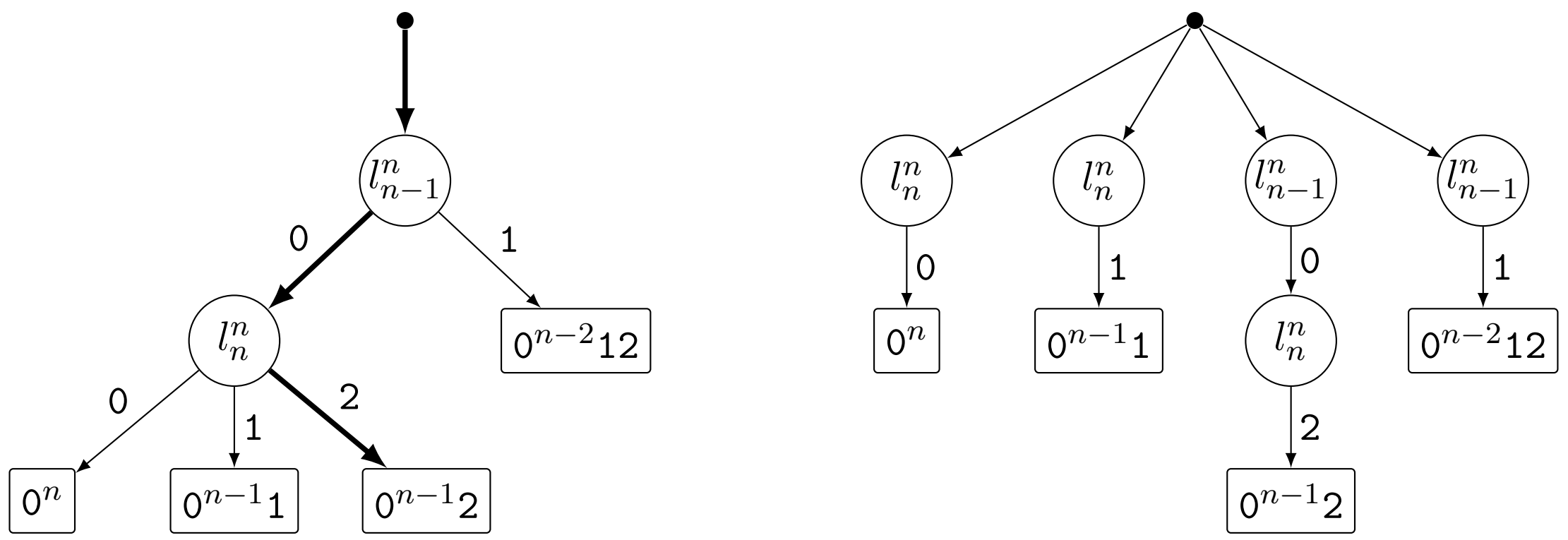

The nondeterministic depths can be described by certificates. This is the standard certificate characterization of nondeterministic query complexity (see \cite{BuhrmanDeWolf2002,Moshkov2020}). For $w \in L(n)$, a set $B \subseteq \{1, \ldots, n\}$ is a \emph{recognition certificate} of $w$ if every $u \in L(n)$ that agrees with $w$ on all positions of $B$ is equal to $w$. For $w \in \Sigma^{n}$, a set $B$ is a \emph{membership certificate} of $w$ if every $u \in \Sigma^{n}$ that agrees with $w$ on $B$ satisfies $u \in L(n) \Leftrightarrow w \in L(n)$.

\begin{proposition}\label{prop:certificates}
If $L(n) \neq \emptyset$, then $\hra_{L}(n)$ is the maximum, over $w \in L(n)$, of the minimum size of a recognition certificate of $w$, and $\hma_{L}(n)$ is the maximum, over $w \in \Sigma^{n}$, of the minimum size of a membership certificate of $w$.
\end{proposition}

\begin{proof}
We consider recognition; membership is analogous. Let $\Gamma$ solve the problem of recognition for $L(n)$ nondeterministically, let $w \in L(n)$, and let $\xi$ be a complete path with $w \in \Sigma^{n}(\xi)$. The set $B$ of positions queried along $\xi$ is a recognition certificate of $w$: if $u \in L(n)$ agrees with $w$ on $B$, then $u \in \Sigma^{n}(\xi)$, so $u$ is the label of the terminal node of $\xi$, i.e., $u = w$. Since $\xi$ has at least $\lvert B \rvert$ nodes labeled with attributes (a position may be queried more than once, which only increases the count), $h(\Gamma)$ is at least the minimum certificate size of $w$. Conversely, for each $w \in L(n)$ fix a certificate $B_{w}$ of minimum size and let $\Gamma$ consist of one path $\xi_{w}$ per word $w$: it leaves the root, queries the positions of $B_{w}$ with edges labeled by the corresponding letters of $w$, and ends in a terminal node labeled with $w$. Condition~2 of Definition~\ref{def:recognition} holds since $w \in \Sigma^{n}(\xi_{w})$. For condition~3, if $w \in \Sigma^{n}(\xi_{w'})$ for some $w' \in L(n)$, then $w$ agrees with $w'$ on the certificate $B_{w'}$ of $w'$, so $w = w'$, which is the label of the terminal node of $\xi_{w'}$. Hence $\Gamma$ solves the problem of recognition for $L(n)$ nondeterministically, and $h(\Gamma) = \max_{w} \lvert B_{w} \rvert$.
\end{proof}

\begin{remark}
The tree constructed in the proof of Proposition~\ref{prop:certificates} has one path per word, so its size is of order $\lvert L(n) \rvert \cdot h(\Gamma)$ for recognition and $k^{n} \cdot h(\Gamma)$ for membership. Since only depth is measured in this paper, a nondeterministic tree of small depth may have a large number of nodes. A deterministic tree of depth $h$ has at most $k^{h}$ terminal nodes, which is polynomial in $n$ when $h = O(\log n)$.
\end{remark}

\section{Main results}\label{sec:results}

Throughout this section, $L$ is a subword-closed language over $\Sigma$ with $\lvert \Sigma \rvert = k \geq 2$. For $k = 2$ and $\Stra \in \{\Strzero, \Strone\}$, Moshkov~\cite{Moshkov2023} uses the word $\Stra^{m}\bar{\Stra}\Stra^{m}$, where $\bar{\Stra}$ denotes the letter different from~$\Stra$. For $k \geq 3$ there is no distinguished ``other letter'', and we quantify over all pairs of different letters. All results of this section are proved in Section~\ref{sec:proofs}: Theorems~\ref{thm:recognition} and~\ref{thm:membership} and Proposition~\ref{prop:classes} in Section~\ref{sec:proofs-main}, using the lemmas of Sections~\ref{sec:structure} and~\ref{sec:upper}, and Propositions~\ref{prop:sorted}, \ref{prop:hom0}, and~\ref{prop:exact}--\ref{prop:effective} in Section~\ref{sec:proofs-examples}.

\begin{definition}[Homogeneity dimension]\label{def:hom}
The \emph{homogeneity set} of $L$ is
\[
    \Shom(L) \;=\; \bigl\{\, m \in \mathbb{N} \;\bigm|\; \exists\, \Stra, \Strb \in \Sigma,\ \Stra \neq \Strb,\ \Stra^{m}\Strb\Stra^{m} \in L \,\bigr\},
\]
and the \emph{homogeneity dimension} of $L$ is
\[
    \Hom(L) \;=\;
    \begin{cases}
        \infty,         & \text{if } \Shom(L) \text{ is infinite}, \\
        \max \Shom(L),  & \text{if } \Shom(L) \text{ is finite and nonempty}, \\
        0,              & \text{if } \Shom(L) = \emptyset.
    \end{cases}
\]
\end{definition}

\begin{definition}[Heterogeneity dimension]\label{def:het}
The \emph{heterogeneity set} of $L$ is
\[
    \Shet(L) \;=\; \bigl\{\, m \in \mathbb{N} \;\bigm|\; \exists\, \Stra, \Strb \in \Sigma,\ \Stra \neq \Strb,\ \Stra^{m}\Strb^{m} \in L \,\bigr\},
\]
and the \emph{heterogeneity dimension} of $L$ is
\[
    \Het(L) \;=\;
    \begin{cases}
        \infty,         & \text{if } \Shet(L) \text{ is infinite}, \\
        \max \Shet(L),  & \text{if } \Shet(L) \text{ is finite and nonempty}, \\
        0,              & \text{if } \Shet(L) = \emptyset.
    \end{cases}
\]
\end{definition}

The value $\Hom(L) = 0$ covers both $\Shom(L) = \emptyset$ and $\Shom(L) = \{0\}$, and similarly for $\Het(L)$; this causes no problem, since the arguments below use only the fact that $\Stra^{m+1}\Strb\Stra^{m+1} \notin L$ when $\Hom(L) = m$ (and $\Stra^{h+1}\Strb^{h+1} \notin L$ when $\Het(L) = h$).

\begin{remark}\label{rem:dimensions}
Since $L$ is subword closed, $\Shom(L)$ and $\Shet(L)$ are downward closed: $\Stra^{m}\Strb\Stra^{m} \in L$ implies $\Stra^{m'}\Strb\Stra^{m'} \in L$, and $\Stra^{m}\Strb^{m} \in L$ implies $\Stra^{m'}\Strb^{m'} \in L$, for all $m' \leq m$. Hence each of these sets is $\mathbb{N}$, an initial segment $\{0, \ldots, M\}$, or $\emptyset$. In particular, if $\Hom(L) = m < \infty$, then $\Stra^{m+1}\Strb\Stra^{m+1} \notin L$ for all $\Stra \neq \Strb$, and since $L$ is subword closed, $\Stra^{m+1}\Strb\Stra^{m+1}$ is not a subword of any word of $L$; the same holds for $\Stra^{h+1}\Strb^{h+1}$ if $\Het(L) = h < \infty$. If $\Hom(L) = \infty$, then a single pair $\Stra \neq \Strb$ satisfies $\Stra^{m}\Strb\Stra^{m} \in L$ for all $m$: for each $m$ some pair works, there are only $k(k - 1)$ pairs, so some pair works for infinitely many $m$, and by downward closure for all $m$. The same holds for $\Het(L) = \infty$. For $k = 2$, the condition $\Stra \neq \Strb$ forces $\Strb = \bar{\Stra}$, and Definitions~\ref{def:hom} and~\ref{def:het} coincide with those of \cite{Moshkov2023}.
\end{remark}

\begin{theorem}\label{thm:recognition}
Let $L$ be a subword-closed language over $\Sigma$, where $\lvert \Sigma \rvert = k \geq 2$, and let $m = \Hom(L)$ and $h = \Het(L)$.
\begin{enumerate}
    \item[(a)] If $m = \infty$, then $\hrd_{L}(n) = \hra_{L}(n) = n$ for every $n$. Hence $\hrd_{L}(n) = \Theta(n)$ and $\hra_{L}(n) = \Theta(n)$.
    \item[(b)] If $m < \infty$ and $h = \infty$, then for every $n$
    \begin{align*}
        \log_{k}(n + 1) \;\leq\; \hrd_{L}(n) &\;\leq\; (2m + 1) \cdot \bigl(2 + 10k\lceil \log_{2} n \rceil\bigr), \\
        \hra_{L}(n) &\;\leq\; k(4m + 3).
    \end{align*}
    Hence $\hrd_{L}(n) = \Theta(\log n)$ and $\hra_{L}(n) = O(1)$.
    \item[(c)] If $m < \infty$ and $h < \infty$, then for every $n$
    \[
        \hra_{L}(n) \;\leq\; \hrd_{L}(n) \;\leq\; 2T + 2, \qquad \text{where } T = k(2m + 1) + (k - 1)h.
    \]
    Hence $\hrd_{L}(n) = O(1)$ and $\hra_{L}(n) = O(1)$.
\end{enumerate}
\end{theorem}

\begin{theorem}\label{thm:membership}
Let $L$ be a subword-closed language over $\Sigma$, where $\lvert \Sigma \rvert = k \geq 2$.
\begin{enumerate}
    \item[(a)] If $\lvert L \rvert = \infty$ and $L^{C} \neq \emptyset$, let $w_{0}$ be a word of minimum length in $L^{C}$. Then $n - \lvert w_{0} \rvert + 1 \leq \hma_{L}(n) \leq \hmd_{L}(n) \leq n$ for every $n \geq \lvert w_{0} \rvert$. Hence $\hmd_{L}(n) = \Theta(n)$ and $\hma_{L}(n) = \Theta(n)$.
    \item[(b)] If $\lvert L \rvert < \infty$, then $\hmd_{L}(n) = \hma_{L}(n) = 0$ for all sufficiently large $n$; if $L^{C} = \emptyset$, i.e., $L = \Sigma^{*}$, then $\hmd_{L}(n) = \hma_{L}(n) = 0$ for every $n$. In both cases $\hmd_{L}(n) = O(1)$ and $\hma_{L}(n) = O(1)$.
\end{enumerate}
\end{theorem}

\begin{remark}\label{rem:bounds}
The bounds in Theorem~\ref{thm:recognition}(b) and~(c) are taken verbatim from Lemmas~\ref{lem:certificate}, \ref{lem:logdepth}, and~\ref{lem:bounded-depth}, where they are derived; the constants are not optimized (see Proposition~\ref{prop:hom0} for a smaller bound when $\Hom(L) = 0$). The two logarithms in Theorem~\ref{thm:recognition}(b) have different bases. The lower bound counts terminal nodes: $L(n)$ contains the $n + 1$ words $\Stra^{i}\Strb^{n-i}$ for suitable letters $\Stra \neq \Strb$, and a deterministic tree of depth $h$ has at most $k^{h}$ terminal nodes. The upper bound comes from a halving procedure. The constant of the logarithmic term must grow with $k$ (Proposition~\ref{prop:sorted}). In case~(c) the bound must depend on $\Het(L)$ as well as on $k$ and $\Hom(L)$: for $L = \mathord{\downarrow}\{\Strzero^{h}\Strone^{h}\} = \{\Strzero^{a}\Strone^{b} : a, b \leq h\}$ we have $\Hom(L) = 0$ and $\Het(L) = h$, while $\lvert L(h) \rvert = h + 1$ and hence $\hrd_{L}(h) \geq \log_{k}(h + 1)$. The bound $\hra_{L}(n) \leq k(4m + 3)$ of Lemma~\ref{lem:certificate} holds whenever $m = \Hom(L) < \infty$, so it also holds in case~(c).
\end{remark}

\begin{example}\label{ex:running}
Let $\Sigma = \{\Strzero, \Strone, \Strtwo\}$ and $L_{3} = \{\Strzero^{i}\Strone^{j}\Strtwo^{l} : i, j, l \in \mathbb{N}\}$, the language of \emph{sorted} ternary words. This language is subword closed, since every subword of $\Strzero^{i}\Strone^{j}\Strtwo^{l}$ has the form $\Strzero^{i'}\Strone^{j'}\Strtwo^{l'}$. No word of $L_{3}$ contains $\Stra\Strb\Stra$ with $\Stra \neq \Strb$ as a subword, while every single letter belongs to $L_{3}$; hence $\Shom(L_{3}) = \{0\}$ and $\Hom(L_{3}) = 0$. Since $\Strzero^{m}\Strone^{m} \in L_{3}$ for all $m$, $\Het(L_{3}) = \infty$. By Theorem~\ref{thm:recognition}(b), $\hrd_{L_{3}}(n) = \Theta(\log n)$ and $\hra_{L_{3}}(n) \leq 9$; the exact values are $\hra_{L_{3}}(n) = \min(n, 4)$ and, by Proposition~\ref{prop:recursion}, $\hrd_{L_{3}}(n) = \lfloor \log_{2} n \rfloor + \lfloor \log_{2}(n/3) \rfloor + 2$ for $n \geq 2$. Since $L_{3}$ is infinite and $\Strone\Strzero \notin L_{3}$, Theorem~\ref{thm:membership}(a) gives $\hmd_{L_{3}}(n) = \Theta(n)$ and $\hma_{L_{3}}(n) = \Theta(n)$; in fact both are equal to $n$ for $n \geq 2$ (Proposition~\ref{prop:exact}). Membership in $L_{3}$ is the problem of testing whether a ternary string is sorted, a problem studied in property testing~\cite{ErgunEtAl2000}. In the exact (zero-error) setting considered here, the whole input has to be read.
\end{example}

\begin{example}\label{ex:ambient}
The membership behavior depends on the alphabet over which $L$ is considered, while $\Hom(L)$ and $\Het(L)$ depend only on the letters occurring in $L$. The language $\{\Strzero, \Strone\}^{*}$ is subword closed both over $\{\Strzero, \Strone\}$ and over $\{\Strzero, \Strone, \Strtwo\}$. Over $\{\Strzero, \Strone\}$ its complement is empty, so the membership problem is trivial: $\hmd(n) = \hma(n) = 0$. Over $\{\Strzero, \Strone, \Strtwo\}$ its complement contains $\Strtwo$, and both membership depths are equal to $n$: a membership certificate of the word $\Strzero^{n} \in L$ must contain every position $p$, since replacing the letter at position $p$ by $\Strtwo$ produces a word outside~$L$.
\end{example}

\begin{proposition}\label{prop:sorted}
Let $\Sigma = \{\Stra_{1}, \ldots, \Stra_{k}\}$ and $\Lsort = \{\Stra_{1}^{i_{1}}\Stra_{2}^{i_{2}} \cdots \Stra_{k}^{i_{k}} : i_{1}, \ldots, i_{k} \in \mathbb{N}\}$. Then $\Lsort$ is subword closed, $\Hom(\Lsort) = 0$, $\Het(\Lsort) = \infty$, and for every $n$
\[
    (k - 1)\log_{k}\frac{n}{k - 1} \;\leq\; \hrd_{\Lsort}(n) \;\leq\; (k - 1)\lceil \log_{2}(n + 1) \rceil.
\]
Moreover, $\hra_{\Lsort}(n) = n$ for $n \leq k$ and $\hra_{\Lsort}(n) = 2(k - 1)$ for $n \geq 2(k - 1)$. Finally, the words $\Stra_{1}^{i}\Stra_{2}^{i} \cdots \Stra_{k}^{i}$ with $i \geq 2$ require $r = k$ long blocks in Lemma~\ref{lem:structure}.
\end{proposition}

The lower bound is the counting bound $\log_{k}\lvert \Lsort(n) \rvert$ combined with the estimate $\lvert \Lsort(n) \rvert = \binom{n + k - 1}{k - 1} \geq \bigl(\frac{n}{k - 1}\bigr)^{k-1}$, and the upper bound comes from $k - 1$ binary searches. Thus, measured in units of $\log_{2} n$, the constant of the logarithmic term for $\Lsort$ lies between $(k - 1)/\log_{2} k$ and $k - 1$; in particular, it is unbounded as $k \to \infty$. For $k = 3$, $\Lsort = L_{3}$, and Proposition~\ref{prop:recursion} shows that $\hrd_{L_{3}}(n) = 2\log_{2} n + O(1)$. Thus, for $k = 3$, the constant of the upper bound is exact, and the counting bound is not tight. We conjecture that $\hrd_{\Lsort}(n) = (k - 1)\log_{2} n + O(k)$ for every $k$. The value of $\hra_{\Lsort}(n)$ for $k < n < 2(k - 1)$, a nonempty range only for $k \geq 4$, remains open.

The procedure behind the logarithmic upper bound gives a smaller bound when $\Hom(L) = 0$, i.e., when no word $\Stra\Strb\Stra$ with $\Stra \neq \Strb$ belongs to $L$.

\begin{proposition}\label{prop:hom0}
If $\Hom(L) = 0$, then $\hrd_{L}(n) \leq 2 + (k - 1)\lceil \log_{2} n \rceil$ for every $n$.
\end{proposition}

For $L_{3}$, Proposition~\ref{prop:hom0} gives $2 + 2\lceil \log_{2} n \rceil$, which exceeds the exact value by at most $4$ (by Proposition~\ref{prop:recursion}), whereas Lemma~\ref{lem:logdepth} gives $2 + 30\lceil \log_{2} n \rceil$.

\subsection{Complexity classes}

To describe the joint behavior of the four functions $\hrd_{L}$, $\hra_{L}$, $\hmd_{L}$, and $\hma_{L}$, consider the five classes $\mathcal{C}_{1}, \ldots, \mathcal{C}_{5}$ of subword-closed languages over $\Sigma$ described in columns 2--5 of Table~\ref{tab:classes}; an empty cell means that no condition is imposed. For example, $\mathcal{C}_{1}$ consists of all subword-closed languages $L$ over $\Sigma$ with $\Hom(L) = \infty$ and $L^{C} \neq \emptyset$.

\begin{table}[ht]
\centering
\caption{Joint behavior of $\hrd_{L}$, $\hra_{L}$, $\hmd_{L}$, and $\hma_{L}$ for subword-closed languages over an alphabet with $k \geq 2$ letters.}
\label{tab:classes}
\begin{tblr}{colspec = {c c c c c | c c c c}, row{1} = {font=\bfseries}, hline{1,Z} = {0.08em}, hline{2} = {0.05em}}
Class & $\Hom(L)$ & $\Het(L)$ & $\lvert L \rvert$ & $L^{C}$ & $\hrd_{L}$ & $\hra_{L}$ & $\hmd_{L}$ & $\hma_{L}$ \\
$\mathcal{C}_{1}$ & $= \infty$ & & & $\neq \emptyset$ & $\Theta(n)$ & $\Theta(n)$ & $\Theta(n)$ & $\Theta(n)$ \\
$\mathcal{C}_{2}$ & $= \infty$ & & & $= \emptyset$ & $\Theta(n)$ & $\Theta(n)$ & $O(1)$ & $O(1)$ \\
$\mathcal{C}_{3}$ & $< \infty$ & $= \infty$ & & & $\Theta(\log n)$ & $O(1)$ & $\Theta(n)$ & $\Theta(n)$ \\
$\mathcal{C}_{4}$ & $< \infty$ & $< \infty$ & $= \infty$ & & $O(1)$ & $O(1)$ & $\Theta(n)$ & $\Theta(n)$ \\
$\mathcal{C}_{5}$ & $< \infty$ & $< \infty$ & $< \infty$ & & $O(1)$ & $O(1)$ & $O(1)$ & $O(1)$ \\
\end{tblr}
\end{table}

\begin{proposition}\label{prop:classes}
For every $k \geq 2$, the classes $\mathcal{C}_{1}, \ldots, \mathcal{C}_{5}$ are pairwise disjoint, every subword-closed language over $\Sigma$ belongs to exactly one of them, and the behavior of $\hrd_{L}$, $\hra_{L}$, $\hmd_{L}$, and $\hma_{L}$ for $L \in \mathcal{C}_{i}$ is as described in Table~\ref{tab:classes}. Moreover, each class is nonempty, $\mathcal{C}_{2} = \{\Sigma^{*}\}$, $\mathcal{C}_{5}$ is the class of all finite subword-closed languages, and $\mathcal{C}_{4}$ is the class of all infinite subword-closed languages $L$ for which $\lvert L(n) \rvert$ is bounded.
\end{proposition}

Table~\ref{tab:examples} lists, for $\Sigma = \{\Strzero, \Strone, \Strtwo\}$, a subword-closed language in each class. The language $L_{1}$ is the ternary counterpart of Moshkov's example $\{\Strzero^{i}\Strone\Strzero^{j}, \Strzero^{i} : i, j \in \mathbb{N}\}$; the language $L_{1}'$ is included because it lies in $\mathcal{C}_{1}$ over $\{\Strzero, \Strone, \Strtwo\}$ but in $\mathcal{C}_{2}$ over $\{\Strzero, \Strone\}$ (Example~\ref{ex:ambient}); the languages $L_{3}$, $L_{4}$, and $L_{5}$ use all three letters. We have $L_{4} = \mathord{\downarrow}\{\Strzero^{i}\Strone\Strtwo : i \in \mathbb{N}\} = \{\Strzero^{i}, \Strzero^{i}\Strone, \Strzero^{i}\Strtwo, \Strzero^{i}\Strone\Strtwo : i \in \mathbb{N}\}$ and $L_{5} = \mathord{\downarrow}\{\Strzero\Strone\Strtwo\} = \{\varepsilon, \Strzero, \Strone, \Strtwo, \Strzero\Strone, \Strzero\Strtwo, \Strone\Strtwo, \Strzero\Strone\Strtwo\}$.

\begin{table}[ht]
\centering
\caption{Examples of subword-closed languages over $\Sigma = \{\Strzero, \Strone, \Strtwo\}$ in each complexity class.}
\label{tab:examples}
\begin{tblr}{colspec = {c l c c c c c}, row{1} = {font=\bfseries}, hline{1,Z} = {0.08em}, hline{2} = {0.05em}}
Name & Language & $\Hom$ & $\Het$ & $\lvert L \rvert$ & $L^{C}$ & Class \\
$L_{1}$ & $\{\Strzero^{i}\Strone\Strzero^{j}, \Strzero^{i} : i, j \in \mathbb{N}\}$ & $\infty$ & $1$ & $\infty$ & $\ni \Strtwo$ & $\mathcal{C}_{1}$ \\
$L_{1}'$ & $\{\Strzero, \Strone\}^{*}$ & $\infty$ & $\infty$ & $\infty$ & $\ni \Strtwo$ & $\mathcal{C}_{1}$ \\
$L_{2}$ & $\Sigma^{*}$ & $\infty$ & $\infty$ & $\infty$ & $= \emptyset$ & $\mathcal{C}_{2}$ \\
$L_{3}$ & $\{\Strzero^{i}\Strone^{j}\Strtwo^{l} : i, j, l \in \mathbb{N}\}$ & $0$ & $\infty$ & $\infty$ & $\ni \Strone\Strzero$ & $\mathcal{C}_{3}$ \\
$L_{4}$ & $\mathord{\downarrow}\{\Strzero^{i}\Strone\Strtwo : i \in \mathbb{N}\}$ & $0$ & $1$ & $\infty$ & $\ni \Strone\Strzero$ & $\mathcal{C}_{4}$ \\
$L_{5}$ & $\mathord{\downarrow}\{\Strzero\Strone\Strtwo\}$ & $0$ & $1$ & $8$ & $\ni \Strone\Strzero$ & $\mathcal{C}_{5}$ \\
\end{tblr}
\end{table}

\subsection{Exact values for the examples}

For the languages of Table~\ref{tab:examples}, all four depths can be determined exactly for every $n$.

\begin{proposition}\label{prop:exact}
For the languages of Table~\ref{tab:examples} and every $n \geq 2$, the minimum depths are as given in Table~\ref{tab:exact}. For $n = 1$, all recognition depths are equal to $1$, and the membership depths are equal to $1$ for $L_{1}$ and $L_{1}'$ and to $0$ for the other languages.
\end{proposition}

\begin{table}[ht]
\centering
\caption{Exact minimum depths for the languages of Table~\ref{tab:examples}, valid for every $n \geq 2$; $f(n)$ is given by Proposition~\ref{prop:recursion}.}
\label{tab:exact}
\begin{tblr}{colspec = {c c | c c c c}, row{1} = {font=\bfseries}, hline{1,Z} = {0.08em}, hline{2} = {0.05em}}
Language & Class & $\hrd_{L}(n)$ & $\hra_{L}(n)$ & $\hmd_{L}(n)$ & $\hma_{L}(n)$ \\
$L_{1}$, $L_{1}'$ & $\mathcal{C}_{1}$ & $n$ & $n$ & $n$ & $n$ \\
$L_{2}$ & $\mathcal{C}_{2}$ & $n$ & $n$ & $0$ & $0$ \\
$L_{3}$ & $\mathcal{C}_{3}$ & $f(n)$ & $\min(n, 4)$ & $n$ & $n$ \\
$L_{4}$ & $\mathcal{C}_{4}$ & $2$ & $2$ & $n$ & $n$ \\
$L_{5}$ & $\mathcal{C}_{5}$ & {$2$ if $n = 2$\\ $0$ if $n \geq 3$} & {$2$ if $n = 2$\\ $0$ if $n \geq 3$} & {$n$ if $n \leq 3$\\ $0$ if $n \geq 4$} & {$n$ if $n \leq 3$\\ $0$ if $n \geq 4$} \\
\end{tblr}
\end{table}

\begin{proposition}\label{prop:recursion}
For $n \geq 1$, $\hrd_{L_{3}}(n) = f(n)$, where $f(0) = 0$ and
\[
    f(n) \;=\; 1 + \min_{1 \leq p \leq n} \max\bigl\{\, f(n - p),\ f(p - 1),\ \lceil \log_{2} p \rceil + \lceil \log_{2}(n - p + 1) \rceil \,\bigr\}.
\]
Moreover, $f(1) = 1$ and, for every $n \geq 2$,
\[
    f(n) \;=\; \lfloor \log_{2} n \rfloor + \lfloor \log_{2}(n/3) \rfloor + 2 \;=\;
    \begin{cases}
        2e,     & \text{if } 2^{e} \leq n < 3 \cdot 2^{e-1}, \\
        2e + 1, & \text{if } 3 \cdot 2^{e-1} \leq n < 2^{e+1},
    \end{cases}
\]
where $e = \lfloor \log_{2} n \rfloor$. Consequently, $2\log_{2} n - 2\log_{2}(3/2) < \hrd_{L_{3}}(n) \leq 2\log_{2} n$ for every $n \geq 2$.
\end{proposition}

In the recursion, $p$ is the position of the first query; the three terms correspond to the answers $\Strzero$ (the first $p$ letters are $\Strzero$ and the rest is a sorted word of length $n - p$), $\Strtwo$ (the last $n - p + 1$ letters are $\Strtwo$), and $\Strone$ (both boundaries are pinned to either side of $p$ and are located by two independent binary searches). The recursion can be evaluated in $O(n^{2})$ time; Table~\ref{tab:recursion} lists some of its values. The two bounds in the last statement of Proposition~\ref{prop:recursion} are sharp: equality on the right holds for $n = 2^{e}$, and $f(n) - 2\log_{2} n$ approaches $-2\log_{2}(3/2) \approx -1.17$ as $n$ approaches $3 \cdot 2^{e-1}$ from below. The counting lower bound $\log_{3}\lvert L_{3}(n) \rvert$ is only about $1.26\log_{2} n$.

\begin{table}[ht]
\centering
\caption{Values of $f(n) = \hrd_{L_{3}}(n)$ computed from the recursion of Proposition~\ref{prop:recursion}. The first two rows give $n = 1, \ldots, 16$, and the last two rows give selected larger values of $n$. For $n \leq 11$ the values agree with the memoized minimax computation described in Section~\ref{sec:computation}.}
\label{tab:recursion}
\begin{tblr}{colspec = {l | c c c c c c c c c c c c c c c c}, hline{1,3,Z} = {0.08em}, hline{2,4} = {0.05em}}
$n$ & 1 & 2 & 3 & 4 & 5 & 6 & 7 & 8 & 9 & 10 & 11 & 12 & 13 & 14 & 15 & 16 \\
$f(n)$ & 1 & 2 & 3 & 4 & 4 & 5 & 5 & 6 & 6 & 6 & 6 & 7 & 7 & 7 & 7 & 8 \\
$n$ & 23 & 24 & 32 & 64 & 128 & 256 & 512 & 1024 & 2048 & 4096 & & & & & & \\
$f(n)$ & 8 & 9 & 10 & 12 & 14 & 16 & 18 & 20 & 22 & 24 & & & & & & \\
\end{tblr}
\end{table}

\subsection{Computational verification}\label{sec:computation}

All results of this paper are proved in Section~\ref{sec:proofs}. The computations described here do not replace these proofs: they check the exact values on finite ranges of $n$ and test the lemmas on random examples. The programs are written in Python~3 and use only the standard library. We checked all exact values of Tables~\ref{tab:exact} and~\ref{tab:recursion} by computer, for $\Sigma = \{\Strzero, \Strone, \Strtwo\}$. Let $U$ be the set of considered words, i.e., $U = L(n)$ for recognition and $U = \Sigma^{n}$ for membership. The \emph{label} of a word of $U$ is the word itself for recognition and the answer to the question $w \in L(n)$ for membership.

The minimum depth of deterministic decision trees was computed by a minimax recursion with memoization; decision trees are not enumerated. A state of the recursion is a word of length $n$ over $\Sigma \cup \{\ast\}$, where $\ast$ means that the letter at this position is not known yet, so there are at most $(k + 1)^{n}$ states. A state is terminal if all words of $U$ that are consistent with it have the same label. For a nonterminal state, each unknown position is tried as the next query, and the worst answer is taken. Each state is processed in time $O(n \lvert U \rvert)$, and a branch is abandoned as soon as it cannot improve the current minimum.

The minimum depth of nondeterministic decision trees was computed by Proposition~\ref{prop:certificates}. We consider the sets $B \subseteq \{1, \ldots, n\}$ in order of increasing size. For each $B$, the words of $U$ are grouped according to their letters at the positions of $B$, and a word is settled at size $\lvert B \rvert$ if all words of its group have the same label. This takes time $O(2^{n} \cdot n \cdot \lvert U \rvert)$.

Recognition depths were computed for $n \leq 11$ for $L_{3}$ and $L_{4}$ (about $3$ seconds), for $n \leq 8$ for $L_{1}$ and $L_{1}'$, and for $n \leq 6$ for $L_{2}$ and $L_{5}$. Membership depths were computed for $n \leq 8$ for $L_{1}$, $L_{1}'$, $L_{3}$, and $L_{4}$ (about $30$ seconds in total, including the corresponding recognition depths), and for $n \leq 6$ for $L_{2}$ and $L_{5}$ (less than one second). All values coincide with Table~\ref{tab:exact}. The values of $\hrd_{L_{3}}(n)$ for $n \leq 11$ were computed directly, without using the recursion of Proposition~\ref{prop:recursion}, and they coincide with Table~\ref{tab:recursion}. We also evaluated the recursion for $n \leq 12000$ (about $20$ seconds); for $2 \leq n \leq 12000$ its values coincide with the closed form, which Proposition~\ref{prop:recursion} proves for all $n \geq 2$.

In addition, we tested the lemmas of Section~\ref{sec:proofs} on randomly generated subword-closed languages $L = \Av(F)$ over $\{\Strzero, \Strone, \Strtwo\}$. The set $F$ contains all words $\Stra^{m+1}\Strb\Stra^{m+1}$ with $\Stra \neq \Strb$, so that $\Hom(L) \leq m$, where $m$ alternates between $0$ and $1$. Besides these, $F$ contains $0$, $1$, or $2$ further words, each of length chosen uniformly from $\{2, \ldots, 5\}$ and with uniformly random letters. In the first experiment we took $16$ languages and all their words of length at most $11$. For each word we computed the representation of Lemma~\ref{lem:structure} as in its proof and checked that it satisfies the lemma. In the second experiment we took $30$ languages and all their words of length at most $13$ for $m = 0$ and at most $11$ for $m = 1$. In every third language, $F$ also contains all words $\Stra^{h+1}\Strb^{h+1}$ with $\Stra \neq \Strb$, where $h$ is chosen uniformly from $\{0, 1\}$; for these languages $\Het(L)$ is finite. For each word we checked the statements of Lemmas~\ref{lem:shape} and~\ref{lem:certificate}, and of Proposition~\ref{prop:hom0} when $m = 0$, and we simulated the procedure of Lemma~\ref{lem:logdepth} and the queries of Lemma~\ref{lem:bounded-depth}. The certificate of Lemma~\ref{lem:certificate} was compared with all other words of $L(n)$; when $\lvert L(n) \rvert > 400$, this was done for a random $5\%$ of the words. In total, more than $5 \cdot 10^{5}$ words were checked in about two minutes, and no violation of the lemmas was found. The scripts and all experimental data are available from the author on request.

\subsection{Effectivity}

As noted after Definition~\ref{def:av}, every subword-closed language $L$ over $\Sigma$ has the form $L = \Av(F)$ for a finite set $F \subseteq \Sigma^{*}$, for example the set of minimal words of $L^{C}$.

\begin{proposition}\label{prop:effective}
Let $F \subseteq \Sigma^{*}$ be a finite set of words with total length $\lVert F \rVert = \sum_{f \in F} \lvert f \rvert$. Then there is an algorithm that, given $F$, computes $\Hom(\Av(F))$ and $\Het(\Av(F))$, decides whether $\Av(F)$ is finite and whether $\Av(F) = \Sigma^{*}$, and hence finds the complexity class of $\Av(F)$, in time $O(k^{2} \cdot \lVert F \rVert)$. Here the input consists of the alphabet $\Sigma$ and the list of words of $F$, and $F$ does not need to consist of the minimal words of the complement of $\Av(F)$.
\end{proposition}

Since $\lVert F \rVert \leq \lvert F \rvert \cdot \max_{f \in F} \lvert f \rvert$, the running time is also $O(k^{2} \cdot \lvert F \rvert \cdot \max_{f \in F} \lvert f \rvert)$.

If $L$ is given by a deterministic finite automaton with transition monoid $M$, the class of $L$ is also computable. Let $\Stra \neq \Strb$, and let $x, y \in M$ be the transformations of the states induced by $\Stra$ and $\Strb$. The sequence $x^{0}, x^{1}, x^{2}, \ldots$ takes at most $\lvert M \rvert$ different values, so it is ultimately periodic with preperiod and period at most $\lvert M \rvert$, and the same holds for the powers of $y$. The state reached on $\Stra^{m}\Strb\Stra^{m}$ depends only on $x^{m}$, so it is an ultimately periodic function of $m$ with preperiod and period at most $\lvert M \rvert$. The state reached on $\Stra^{m}\Strb^{m}$ depends on the pair $(x^{m}, y^{m})$, so it is ultimately periodic with preperiod at most $\lvert M \rvert$ and period at most $\lvert M \rvert^{2}$, the least common multiple of the two periods. Since $\Shom(L)$ and $\Shet(L)$ are downward closed, it is therefore enough to test finitely many values of $m$. The complexity of this computation is left open (Section~\ref{sec:conclusions}).

\section{Proofs}\label{sec:proofs}

Throughout this section, $L$ is a subword-closed language over $\Sigma$ with $\lvert \Sigma \rvert = k \geq 2$.

\subsection{Structure of words}\label{sec:structure}

\begin{lemma}\label{lem:structure}
Let $m = \Hom(L) < \infty$. Then every word $w \in L$ can be represented in the form
\begin{equation}\label{eq:structure}
    w \;=\; w_{0}\, \Strc_{1}^{i_{1}}\, w_{1}\, \Strc_{2}^{i_{2}}\, w_{2} \cdots \Strc_{r}^{i_{r}}\, w_{r},
\end{equation}
where $0 \leq r \leq k$, the letters $\Strc_{1}, \ldots, \Strc_{r} \in \Sigma$ are pairwise different, $i_{1}, \ldots, i_{r} \in \mathbb{N}$, and $w_{0}, \ldots, w_{r} \in \Sigma^{*}$ satisfy
\[
    \lvert w_{0} \rvert + \cdots + \lvert w_{r} \rvert \;\leq\; 2mr + (2m + 1)(k - r) \;\leq\; t, \qquad \text{where } t = k(2m + 1).
\]
We call the factors $\Strc_{j}^{i_{j}}$ the \emph{blocks} and the factors $w_{j}$ the \emph{gaps} of the representation.
\end{lemma}

\begin{proof}
By Remark~\ref{rem:dimensions}, no word $\Stra^{m+1}\Strb\Stra^{m+1}$ with $\Stra \neq \Strb$ is a subword of a word of $L$. Let $w = a_{1} \cdots a_{n} \in L$. Call a letter $\Stra$ \emph{heavy} if $\lvert w \rvert_{\Stra} \geq 2m + 2$ and \emph{light} otherwise. Let $\Stra$ be heavy, $N = \lvert w \rvert_{\Stra}$, and let $p_{1} < \cdots < p_{N}$ be the positions of the occurrences of $\Stra$ in $w$. Put $\ell_{\Stra} = p_{m+1}$ and $\rho_{\Stra} = p_{N-m}$; since $N \geq 2m + 2$, $\ell_{\Stra} < \rho_{\Stra}$. If $a_{p} = \Strb \neq \Stra$ for some $p$ with $\ell_{\Stra} < p < \rho_{\Stra}$, then the letters at positions $p_{1}, \ldots, p_{m+1}, p, p_{N-m}, \ldots, p_{N}$ form the subword $\Stra^{m+1}\Strb\Stra^{m+1}$ of $w$, which is impossible. Hence the factor $B_{\Stra} = a_{\ell_{\Stra}} \cdots a_{\rho_{\Stra}}$ is a power of $\Stra$, and exactly $2m$ occurrences of $\Stra$ lie outside $B_{\Stra}$, namely at the positions $p_{1}, \ldots, p_{m}$ and $p_{N-m+1}, \ldots, p_{N}$.

Let $\Strc_{1}, \ldots, \Strc_{r}$ be the heavy letters, numbered so that $\ell_{\Strc_{1}} < \cdots < \ell_{\Strc_{r}}$; clearly $r \leq k$. The factors $B_{\Strc_{j}}$ occupy pairwise disjoint intervals of positions, since they consist of different letters; hence $B_{\Strc_{1}}, \ldots, B_{\Strc_{r}}$ occur in $w$ in this order from left to right. Thus $w$ has the form \eqref{eq:structure} with $\Strc_{j}^{i_{j}} = B_{\Strc_{j}}$, where $w_{0}$ is the prefix before $B_{\Strc_{1}}$, $w_{j}$ is the factor between $B_{\Strc_{j}}$ and $B_{\Strc_{j+1}}$, and $w_{r}$ is the suffix after $B_{\Strc_{r}}$ (if $r = 0$, then $w = w_{0}$). To bound the total length of the gaps, we sum over the letters of $\Sigma$ the number of occurrences of each letter outside the blocks: each of the $r$ heavy letters has exactly $2m$ such occurrences, and each of the $k - r$ light letters has at most $2m + 1$ occurrences in total. This gives $\sum_{j} \lvert w_{j} \rvert \leq 2mr + (2m + 1)(k - r) \leq k(2m + 1)$.
\end{proof}

\begin{remark}\label{rem:lemma1fails}
For $k = 2$, Lemma~\ref{lem:structure} gives the form $w_{0}\Stra^{i}w_{1}\bar{\Stra}^{j}w_{2}$ of \cite[Lemma~1]{Moshkov2023}, where Moshkov bounds each gap by $2m$ and hence the total by $6m$, while Lemma~\ref{lem:structure} bounds the total by $4m + 2$; the two bounds are incomparable gap-wise, and the total bound here is smaller for $m \geq 2$. For $k \geq 3$, the binary form with at most two long blocks is no longer valid: in the language $L_{3}$ of Example~\ref{ex:running}, with $\Hom(L_{3}) = 0$, the words $\Strzero^{n}\Strone^{n}\Strtwo^{n}$ cannot be written with at most two powers of letters and gaps of bounded total length. More generally, by Proposition~\ref{prop:sorted}, the number $r = k$ of blocks in Lemma~\ref{lem:structure} is attained for every $k$.
\end{remark}

\begin{lemma}\label{lem:shape}
Let $m = \Hom(L) < \infty$ and $h = \Het(L) < \infty$, and put $T = k(2m + 1) + (k - 1)h$. Then every word $w \in L$ can be written as $w = u\,\Strc^{i}\,v$ with $\Strc \in \Sigma$, $i \in \mathbb{N}$, and $u, v \in \Sigma^{*}$ such that $\lvert u \rvert + \lvert v \rvert \leq T$.
\end{lemma}

\begin{proof}
Represent $w$ in the form \eqref{eq:structure}. If $r = 0$, then $\lvert w \rvert \leq t \leq T$ and we take $u = w$, $i = 0$, and $v = \varepsilon$. Let $r \geq 1$. We first show that at most one of the exponents $i_{1}, \ldots, i_{r}$ exceeds $h$. Let $1 \leq j < j' \leq r$ and $s = \min(i_{j}, i_{j'})$. In the proof of Lemma~\ref{lem:structure} the blocks are numbered by increasing $\ell_{\Strc_{j}}$, i.e., in the order in which they occur in $w$, so the block $\Strc_{j}^{i_{j}}$ lies entirely to the left of the block $\Strc_{j'}^{i_{j'}}$. Choosing $s$ occurrences of $\Strc_{j}$ in the first block, followed from left to right by~$s$ occurrences of $\Strc_{j'}$ in the second block, we see that $\Strc_{j}^{s}\Strc_{j'}^{s}$ is a subword of $w$ (the gaps between the two blocks are simply deleted). Since $L$ is subword closed, $\Strc_{j}^{s}\Strc_{j'}^{s} \in L$, and since $\Strc_{j} \neq \Strc_{j'}$, we get $s \in \Shet(L)$. By Remark~\ref{rem:dimensions}, $\Shet(L) \subseteq \{0, \ldots, h\}$, so $s \leq h$. Thus $\min(i_{j}, i_{j'}) \leq h$ for every pair $j < j'$; if two exponents exceeded $h$, their minimum would exceed $h$, which is impossible. Hence at most one exponent exceeds $h$.

Let $j^{*}$ be an index with $i_{j^{*}} = \max_{j} i_{j}$, so that $i_{j} \leq h$ for all $j \neq j^{*}$. Let $u$ be the prefix of $w$ before the block $\Strc_{j^{*}}^{i_{j^{*}}}$, let $v$ be the suffix after it, and put $\Strc = \Strc_{j^{*}}$ and $i = i_{j^{*}}$. The words $u$ and $v$ consist of all gaps $w_{0}, \ldots, w_{r}$ and all blocks other than the $j^{*}$-th one, so $\lvert u \rvert + \lvert v \rvert \leq t + (r - 1)h \leq k(2m + 1) + (k - 1)h = T$.
\end{proof}

\begin{corollary}\label{cor:bounded}
Let $\Hom(L) < \infty$ and $\Het(L) < \infty$, and let $T$ be as in Lemma~\ref{lem:shape}. Then $\lvert L(n) \rvert \leq p$ for every $n$, where $p = (T + 1)k^{T+2}$.
\end{corollary}

\begin{proof}
By Lemma~\ref{lem:shape}, every $w \in L(n)$ can be written as $u\Strc^{i}v$ with $\lvert u \rvert + \lvert v \rvert \leq T$, and $w$ is determined by $n$ and the triple $(u, \Strc, v)$, since $i = n - \lvert u \rvert - \lvert v \rvert$. The number of pairs $(u, v)$ with $\lvert u \rvert + \lvert v \rvert = \ell$ is $(\ell + 1)k^{\ell}$, so the number of triples is at most $k\sum_{\ell = 0}^{T}(\ell + 1)k^{\ell} \leq k(T + 1)\sum_{\ell = 0}^{T} k^{\ell} < (T + 1)k^{T+2}$.
\end{proof}

\begin{remark}\label{rem:converse}
The converse of Corollary~\ref{cor:bounded} also holds: if $\Hom(L) = \infty$, then by Remark~\ref{rem:dimensions} the set $L(n)$ contains the $n$ pairwise different words $\Stra^{i}\Strb\Stra^{n-1-i}$, $i = 0, \ldots, n - 1$, and if $\Het(L) = \infty$, then $L(n)$ contains the $n + 1$ pairwise different words $\Stra^{i}\Strb^{n-i}$, $i = 0, \ldots, n$. Thus $\lvert L(n) \rvert$ is bounded if and only if $\Hom(L) < \infty$ and $\Het(L) < \infty$.
\end{remark}

\subsection{Upper bounds for recognition}\label{sec:upper}

\begin{lemma}\label{lem:certificate}
Let $m = \Hom(L) < \infty$. Then $\hra_{L}(n) \leq k(4m + 3)$ for every $n$.
\end{lemma}

\begin{proof}
Let $L(n) \neq \emptyset$ and $w = a_{1} \cdots a_{n} \in L(n)$, and fix a representation \eqref{eq:structure} of $w$. Let $B_{w}$ consist of all positions of the gaps $w_{0}, \ldots, w_{r}$, and, for each block $\Strc_{j}^{i_{j}}$ occupying the positions $x, \ldots, y$, of all these positions if $i_{j} \leq 2m + 1$, and of the positions $x, \ldots, x + m$ and $y - m, \ldots, y$ otherwise. Then $\lvert B_{w} \rvert \leq t + r(2m + 2) \leq k(2m + 1) + k(2m + 2) = k(4m + 3)$. Let $u = b_{1} \cdots b_{n} \in L(n)$ agree with $w$ on $B_{w}$. Then $u$ agrees with $w$ on all gaps and all short blocks. Consider a long block $\Stra^{y-x+1}$ occupying the positions $x, \ldots, y$. Then $b_{x} = \cdots = b_{x+m} = \Stra = b_{y-m} = \cdots = b_{y}$, and if $b_{p} = \Strb \neq \Stra$ for some $x + m < p < y - m$, then $\Stra^{m+1}\Strb\Stra^{m+1}$ would be a subword of $u \in L$, which is impossible by Remark~\ref{rem:dimensions}. Hence $u = w$, i.e., $B_{w}$ is a recognition certificate of $w$, and the claim follows from Proposition~\ref{prop:certificates}.
\end{proof}

\begin{lemma}\label{lem:logdepth}
Let $m = \Hom(L) < \infty$. Then $\hrd_{L}(n) \leq (2m + 1)\bigl(2 + 10k\lceil \log_{2} n \rceil\bigr)$ for every $n$.
\end{lemma}

\begin{proof}
Let $L(n) \neq \emptyset$ and $s = m + 1$. If $n < 2s$, the deterministic tree that computes $l^{n}_{1}, \ldots, l^{n}_{n}$ one after another solves the problem with $n \leq 2m + 1$ queries. Let $n \geq 2s$ and $q = \lfloor n/s \rfloor \geq 2$. Partition $\{1, \ldots, n\}$ into consecutive intervals $S_{1}, \ldots, S_{q}$ (\emph{segments}), where $S_{1}, \ldots, S_{q-1}$ have length $s$ and $S_{q}$ has length $n - (q - 1)s \in \{s, \ldots, 2s - 1\}$. To \emph{query a segment} means to compute all attributes $l^{n}_{p}$ with $p$ in that segment. Let $w = a_{1} \cdots a_{n} \in L(n)$. A segment $S_{\gamma}$ is \emph{$\Stra$-constant} if $a_{p} = \Stra$ for all $p \in S_{\gamma}$. We use two facts.
\begin{enumerate}
    \item If $\alpha < \beta$ and $S_{\alpha}$ and $S_{\beta}$ are both $\Stra$-constant, then so is every $S_{\gamma}$ with $\alpha < \gamma < \beta$. Indeed, each position $p$ strictly between $S_{\alpha}$ and $S_{\beta}$ has at least $m + 1$ occurrences of $\Stra$ on each side, so $a_{p} = \Strb \neq \Stra$ would make $\Stra^{m+1}\Strb\Stra^{m+1}$ a subword of $w$.
    \item Call $\gamma \in \{1, \ldots, q - 1\}$ a \emph{change index} of $w$ if the factor of $w$ occupying $S_{\gamma} \cup S_{\gamma+1}$ is not a power of a single letter. Then $w$ has at most $10k$ change indices. Indeed, fix a representation \eqref{eq:structure} of $w$. A change index $\gamma$ yields two adjacent positions $p, p + 1$ in $S_{\gamma} \cup S_{\gamma+1}$ with $a_{p} \neq a_{p+1}$. Since each block is a power of a single letter, either one of $p, p + 1$ belongs to a gap, or $p$ is the last position of a block and $p + 1$ the first position of the next block, the gap between them being empty; there are at most $k - 1$ such pairs. Let $G$ be the set of segments that contain a position of a gap or the position $p$ of such a pair. A gap of length $\ell \geq 1$ meets at most $\ell/s + 2$ segments, because all segments have length at least $s$; hence
    \[
        \lvert G \rvert \;\leq\; \frac{t}{s} + 2(r + 1) + (k - 1) \;<\; 2k + 2(k + 1) + k - 1 \;=\; 5k + 1,
    \]
    using $t/s = k(2m + 1)/(m + 1) < 2k$ and $r \leq k$. Since $\lvert G \rvert$ is an integer, $\lvert G \rvert < 5k + 1$ implies $\lvert G \rvert \leq 5k$. For every change index $\gamma$, at least one of $S_{\gamma}$ and $S_{\gamma+1}$ belongs to $G$, and each segment of $G$ is $S_{\gamma}$ or $S_{\gamma+1}$ for at most two values of $\gamma$; so there are at most $2\lvert G \rvert \leq 10k$ change indices.
\end{enumerate}
The tree implements an adaptive procedure built from the following recursive subroutine. For integers $\alpha, \beta$ with $1 \leq \alpha < \beta \leq q$ such that the segments $S_{\alpha}$ and $S_{\beta}$ have already been queried, $\textsc{Fill}(\alpha, \beta)$ does the following:
\begin{enumerate}
    \item[(F-i)] if $\beta = \alpha + 1$, it stops;
    \item[(F-ii)] if there is a letter $\Stra$ such that both $S_{\alpha}$ and $S_{\beta}$ are $\Stra$-constant, it stops;
    \item[(F-iii)] otherwise, it queries the segment $S_{\gamma}$ with $\gamma = \lfloor (\alpha + \beta)/2 \rfloor$, then calls $\textsc{Fill}(\alpha, \gamma)$, and after that call has stopped, calls $\textsc{Fill}(\gamma, \beta)$.
\end{enumerate}
In step (F-iii) we have $\beta \geq \alpha + 2$ and hence $\alpha < \gamma < \beta$, so both recursive calls satisfy the input conditions; since $\beta - \alpha$ decreases strictly, the recursion terminates. Whether a step applies is decided by answers already obtained, so $\textsc{Fill}$ is a well-defined adaptive query procedure. The \emph{recursion depth} of a call is $0$ for the initial call and $d + 1$ for a call made by a call of depth~$d$. The procedure queries $S_{1}$ and $S_{q}$ (recall that $q \geq 2$) and then calls $\textsc{Fill}(1, q)$. When it ends, every segment has been queried or lies between two queried $\Stra$-constant segments and is therefore $\Stra$-constant by fact~1. Hence the answers determine $w$ uniquely among the words of $L(n)$. Implementing the procedure in the standard way (each node computes the next attribute, the edges leaving it are labeled with the answers that occur for at least one word of $L(n)$ consistent with the previous answers, and each terminal node is labeled with the unique word of $L(n)$ consistent with the answers along its path) yields a deterministic decision tree over $L(n)$ that solves the problem of recognition for $L(n)$ deterministically.

The calls of $\textsc{Fill}$ at the same recursion depth $d$ have intervals $[\alpha, \beta]$ that overlap only in endpoints and satisfy $\beta - \alpha \leq \lceil (q - 1)/2^{d} \rceil$. A call queries a segment only if $\beta - \alpha \geq 2$, which requires $2^{d} < q - 1$; this is possible for at most $\lceil \log_{2} n \rceil$ values of $d$. If $\textsc{Fill}(\alpha, \beta)$ queries a segment, then $\{\alpha, \ldots, \beta - 1\}$ contains a change index of $w$, since otherwise $S_{\alpha} \cup \cdots \cup S_{\beta}$ would carry a power of a single letter $\Stra$ and both $S_{\alpha}$ and $S_{\beta}$ would be $\Stra$-constant. These sets are pairwise disjoint for calls at the same depth, so at most $10k$ segments are queried at each depth. Altogether at most $2 + 10k\lceil \log_{2} n \rceil$ segments are queried, each of at most $2m + 1$ positions, and the bound follows.
\end{proof}

\begin{lemma}\label{lem:bounded-depth}
Let $\Hom(L) < \infty$ and $\Het(L) < \infty$, and let $T$ be as in Lemma~\ref{lem:shape}. Then $\hrd_{L}(n) \leq 2T + 2$ for every $n$.
\end{lemma}

\begin{proof}
Let $L(n) \neq \emptyset$. If $n \leq 2T + 2$, computing all $n$ attributes suffices. Let $n \geq 2T + 3$ and let $Q$ consist of the first $T + 1$ positions $1, \ldots, T + 1$ and the last $T + 1$ positions $n - T, \ldots, n$; these $2T + 2$ positions are distinct. We claim that two words $w, w' \in L(n)$ that agree on $Q$ are equal. By Lemma~\ref{lem:shape}, $w = u\Strc^{i}v$ with $\lvert u \rvert, \lvert v \rvert \leq T$; since $n \geq 2T + 3$, we have $i = n - \lvert u \rvert - \lvert v \rvert \geq 3$, so the block is nonempty, and it occupies the positions $\lvert u \rvert + 1, \ldots, n - \lvert v \rvert$, which include all positions from $T + 1$ to $n - T$. Hence $w$ carries the letter $\Strc$ at all positions $T + 1, \ldots, n - T$, and $\Strc$ is the letter of $w$ at position $T + 1$. The same holds for $w' = u'\Strc'^{i'}v'$. Since $w$ and $w'$ agree at position $T + 1 \in Q$, we get $\Strc = \Strc'$, so $w$ and $w'$ agree at all positions from $T + 1$ to $n - T$, and they agree at all other positions because these belong to $Q$. Thus the deterministic tree that computes the attributes $l^{n}_{p}$, $p \in Q$, one after another (with edges labeled by the answers that occur for words of $L(n)$, and terminal nodes labeled by the unique consistent word) solves the problem of recognition for $L(n)$ deterministically and has depth $2T + 2$.
\end{proof}

\subsection{Proofs of the main theorems}\label{sec:proofs-main}

\begin{proof}[Proof of Theorem~\ref{thm:recognition}]
(a) Let $\Hom(L) = \infty$. By Remark~\ref{rem:dimensions}, there are letters $\Stra \neq \Strb$ with $\Stra^{n}\Strb\Stra^{n} \in L$, so $\Stra^{n} \in L(n)$ and $\Stra^{i}\Strb\Stra^{n-i-1} \in L(n)$ for $i = 0, \ldots, n - 1$. Let $\Gamma$ solve the problem of recognition for $L(n)$ nondeterministically, and let $\xi$ be a complete path with $\Stra^{n} \in \Sigma^{n}(\xi)$; its terminal node is labeled with $\Stra^{n}$. If, for some $i \in \{0, \ldots, n - 1\}$, the attribute $l^{n}_{i+1}$ is not attached to any node of $\xi$, then $\Stra^{i}\Strb\Stra^{n-i-1} \in \Sigma^{n}(\xi)$, which contradicts condition~3 of Definition~\ref{def:recognition}. Hence $h(\Gamma) \geq n$ and $\hra_{L}(n) \geq n$. Since $\hrd_{L}(n) \leq n$, we obtain $\hra_{L}(n) = \hrd_{L}(n) = n$.

(b) Let $\Hom(L) = m < \infty$ and $\Het(L) = \infty$. The upper bounds are Lemmas~\ref{lem:certificate} and~\ref{lem:logdepth}. By Remark~\ref{rem:dimensions}, there are letters $\Stra \neq \Strb$ with $\Stra^{n}\Strb^{n} \in L$, so the $n + 1$ pairwise different words $\Stra^{i}\Strb^{n-i}$, $i = 0, \ldots, n$, belong to $L(n)$. If $\Gamma$ solves the problem of recognition for $L(n)$ deterministically, each of these words labels a terminal node of $\Gamma$ (by conditions~2 and~3 of Definition~\ref{def:recognition}), while $\Gamma$ has at most $k^{h(\Gamma)}$ terminal nodes. Hence $h(\Gamma) \geq \log_{k}(n + 1)$.

(c) This is Lemma~\ref{lem:bounded-depth}, together with $\hra_{L}(n) \leq \hrd_{L}(n)$.
\end{proof}

\begin{proof}[Proof of Theorem~\ref{thm:membership}]
(a) Let $\lvert L \rvert = \infty$ and $L^{C} \neq \emptyset$, and let $w_{0}$ be a word of minimum length in $L^{C}$. Since $L$ is infinite, it contains words of every length (an infinite language over a finite alphabet contains arbitrarily long words, and their prefixes belong to $L$), so $L(n) \neq \emptyset$ for every $n$. Let $n \geq \lvert w_{0} \rvert$ and let $\Gamma$ solve the problem of membership for $L(n)$ nondeterministically. Choose $w \in L(n)$ and a complete path $\xi$ with $w \in \Sigma^{n}(\xi)$; the terminal node of $\xi$ is labeled with $1$. Suppose that at most $n - \lvert w_{0} \rvert$ nodes of $\xi$ are labeled with attributes. Then there are positions $p_{1} < \cdots < p_{\lvert w_{0} \rvert}$ not queried along $\xi$. Let $w'$ be obtained from $w$ by writing the $j$th letter of $w_{0}$ at position $p_{j}$ for $j = 1, \ldots, \lvert w_{0} \rvert$. Then $w' \in \Sigma^{n}(\xi)$, and $w_{0}$ is a subword of $w'$, so $w' \notin L(n)$ because $L$ is subword closed; this contradicts the label $1$ of the terminal node of $\xi$. Hence every such path has at least $n - \lvert w_{0} \rvert + 1$ nodes labeled with attributes, and $\hma_{L}(n) \geq n - \lvert w_{0} \rvert + 1$. The upper bound $\hmd_{L}(n) \leq n$ is given by the deterministic tree that computes $l^{n}_{1}, \ldots, l^{n}_{n}$ one after another, branching on all $k$ letters at each node, with the terminal node reached along $a_{1}, \ldots, a_{n}$ labeled with $1$ if $a_{1} \cdots a_{n} \in L(n)$ and with $0$ otherwise.

(b) If $\lvert L \rvert < \infty$, then $L(n) = \emptyset$, and hence all depths are $0$, for all sufficiently large $n$. If $L^{C} = \emptyset$, then $L = \Sigma^{*}$ and $L(n) = \Sigma^{n}$ for every $n$, so every input belongs to $L(n)$. Hence, for every $n$, the tree consisting of the root, one edge, and a terminal node labeled with $1$ solves the problem of membership for $L(n)$ deterministically and has depth $0$.
\end{proof}

\begin{proof}[Proof of Proposition~\ref{prop:classes}]
The rows of Table~\ref{tab:classes} split first by whether $\Hom(L)$ is finite. If $\Hom(L) = \infty$, the rows $\mathcal{C}_{1}$ and $\mathcal{C}_{2}$ split by whether $L^{C}$ is empty; if $\Hom(L) < \infty$, the rows $\mathcal{C}_{3}$, $\mathcal{C}_{4}$, and $\mathcal{C}_{5}$ split by whether $\Het(L)$ is finite and, if so, whether $L$ is finite. Hence the conditions are mutually exclusive and exhaustive. The entries in the last four columns follow from Theorems~\ref{thm:recognition} and~\ref{thm:membership} and three observations: (i) if $\Hom(L) = \infty$, then $L$ contains $\Stra^{m}\Strb\Stra^{m}$ for all $m$, so $\lvert L \rvert = \infty$; (ii) if $\Het(L) = \infty$, then similarly $\lvert L \rvert = \infty$; and (iii) if $\Hom(L) = m < \infty$, then $\Stra^{m+1}\Strb\Stra^{m+1} \in L^{C}$ for any $\Stra \neq \Strb$, so $L^{C} \neq \emptyset$. For instance, for $L \in \mathcal{C}_{3}$ we have $\lvert L \rvert = \infty$ by (ii) and $L^{C} \neq \emptyset$ by (iii), so Theorem~\ref{thm:membership}(a) applies; for $L \in \mathcal{C}_{1}$, Theorem~\ref{thm:membership}(a) applies by (i).

Since $L^{C} = \emptyset$ means $L = \Sigma^{*}$, and $\Hom(\Sigma^{*}) = \infty$, we have $\mathcal{C}_{2} = \{\Sigma^{*}\}$. If $L$ is finite, then $\Hom(L)$ and $\Het(L)$ are finite by (i) and (ii), so the finite subword-closed languages are exactly those in $\mathcal{C}_{5}$. By Corollary~\ref{cor:bounded} and Remark~\ref{rem:converse}, $\lvert L(n) \rvert$ is bounded if and only if $\Hom(L) < \infty$ and $\Het(L) < \infty$, so $\mathcal{C}_{4}$ consists of the infinite languages with bounded $\lvert L(n) \rvert$. For nonemptiness, if $k \geq 3$, rename three letters of $\Sigma$ as $\Strzero, \Strone, \Strtwo$; the languages of Table~\ref{tab:examples}, with $L_{2}$ replaced by $\Sigma^{*}$, are subword closed over $\Sigma$, and their dimensions, finiteness, and nonempty complements do not change. If $k = 2$, Moshkov's examples $\{\Strzero^{i}\Strone\Strzero^{j}, \Strzero^{i}\}$, $\{\Strzero, \Strone\}^{*}$, $\{\Strzero^{i}\Strone^{j}\}$, $\{\Strzero^{i}\}$, and $\{\Strzero\}$ lie in $\mathcal{C}_{1}, \ldots, \mathcal{C}_{5}$, respectively~\cite{Moshkov2023}.
\end{proof}

\subsection{Proofs for the examples}\label{sec:proofs-examples}

\begin{proof}[Proof of Proposition~\ref{prop:sorted}]
Every subword of a sorted word is sorted, so $\Lsort$ is subword closed. Each single letter belongs to $\Lsort$, while no word $\Stra\Strb\Stra$ with $\Stra \neq \Strb$ is sorted; hence $\Shom(\Lsort) = \{0\}$ and $\Hom(\Lsort) = 0$. Since $\Stra_{1}^{m}\Stra_{2}^{m} \in \Lsort$ for all $m$, $\Het(\Lsort) = \infty$.

A word $w \in \Lsort(n)$ is determined by the numbers $b_{j} = \lvert w \rvert_{\Stra_{1}} + \cdots + \lvert w \rvert_{\Stra_{j}}$, $j = 1, \ldots, k - 1$, with $0 \leq b_{1} \leq \cdots \leq b_{k-1} \leq n$; conversely, every such sequence defines a word of $\Lsort(n)$. Hence $\lvert \Lsort(n) \rvert = \binom{n + k - 1}{k - 1} \geq \bigl(\frac{n}{k - 1}\bigr)^{k-1}$, and the lower bound on $\hrd_{\Lsort}(n)$ follows since a deterministic tree has at most $k^{h}$ terminal nodes. For the upper bound, note that the letter of $w$ at position $p$ is $\Stra_{l}$ with $l \leq j$ if and only if $p \leq b_{j}$. Hence each $b_{j} \in \{0, \ldots, n\}$ can be found by binary search with $\lceil \log_{2}(n + 1) \rceil$ queries, and the $k - 1$ binary searches determine $w$.

For the nondeterministic depth, let $w \in \Lsort(n)$ and let $B$ consist of the positions $b_{j}$ (if $b_{j} \geq 1$) and $b_{j} + 1$ (if $b_{j} \leq n - 1$) for $j = 1, \ldots, k - 1$. If $u \in \Lsort(n)$ agrees with $w$ on $B$, then, with $b_{j}'$ the corresponding numbers of $u$, we have $b_{j}' = b_{j}$ for every $j$. Indeed, if $b_{j} \geq 1$, the letter of $u$ at position $b_{j}$ has index at most $j$, so $b_{j}' \geq b_{j}$; if $b_{j} = 0$, the inequality $b_{j}' \geq b_{j}$ is trivial. Similarly, if $b_{j} \leq n - 1$, the letter of $u$ at position $b_{j} + 1$ has index greater than $j$, so $b_{j}' \leq b_{j}$; if $b_{j} = n$, the inequality $b_{j}' \leq b_{j}$ is trivial. Hence $u = w$, $B$ is a recognition certificate, and $\hra_{\Lsort}(n) \leq \min(n, 2(k - 1))$. For the lower bound when $n \leq k$, consider $w = \Stra_{1}\Stra_{2} \cdots \Stra_{n}$: replacing the letter at position $p < n$ by $\Stra_{p+1}$, or at position $n$ by $\Stra_{n-1}$ (by $\Stra_{2}$ if $n = 1$), gives another word of $\Lsort(n)$, so every recognition certificate of $w$ contains all $n$ positions. When $n \geq 2(k - 1)$, choose $w = \Stra_{1}^{i_{1}} \cdots \Stra_{k}^{i_{k}}$ with $i_{1}, i_{k} \geq 1$ and $i_{2}, \ldots, i_{k-1} \geq 2$. For each $j$, decreasing $b_{j}$ by one (the last letter of the $j$-th block becomes $\Stra_{j+1}$) or increasing it by one (the first letter of the $(j + 1)$-th block becomes $\Stra_{j}$) gives another word of $\Lsort(n)$ that differs from $w$ only at position $b_{j}$, respectively $b_{j} + 1$. These $2(k - 1)$ positions are pairwise different by the choice of the exponents, so every certificate of $w$ contains all of them. Finally, in $\Stra_{1}^{i} \cdots \Stra_{k}^{i}$ with $i \geq 2 = 2\Hom(\Lsort) + 2$, all $k$ letters are heavy in the sense of the proof of Lemma~\ref{lem:structure}.
\end{proof}

\begin{proof}[Proof of Proposition~\ref{prop:hom0}]
If $n = 1$, one query suffices. Let $n \geq 2$. Since $\Hom(L) = 0$, we have $1 \notin \Shom(L)$, so no word $\Stra\Strb\Stra$ with $\Stra \neq \Strb$ is a subword of a word of $L$. Hence, in every $w \in L$, the occurrences of each letter form a single factor, $w$ has at most $k$ runs (maximal factors that are powers of a single letter), and $w$ has at most $k - 1$ positions $\gamma$ with $a_{\gamma} \neq a_{\gamma+1}$. We use the procedure of the proof of Lemma~\ref{lem:logdepth} with $m = 0$, i.e., with segments of length $s = 1$, so that $q = n$ and each segment is a single position. Fact~1 of that proof holds with $m = 0$. A change index is now exactly a position $\gamma$ with $a_{\gamma} \neq a_{\gamma+1}$, so $w$ has at most $k - 1$ change indices, and the counting argument of that proof shows that at most $k - 1$ positions are queried at each of at most $\lceil \log_{2} n \rceil$ recursion depths, in addition to the two initial queries.
\end{proof}

\begin{proof}[Proof of Proposition~\ref{prop:recursion}]
A word of $L_{3}(n)$ is determined by the pair $(a, b) = (\lvert w \rvert_{\Strzero}, \lvert w \rvert_{\Strzero} + \lvert w \rvert_{\Strone})$ with $0 \leq a \leq b \leq n$, and the answer to a query at position $p$ is $\Strzero$ if $p \leq a$, $\Strone$ if $a < p \leq b$, and $\Strtwo$ if $p > b$. The minimum depth of a deterministic decision tree equals the value of the minimax recursion over knowledge states, in which each query is followed by the worst answer. Consider the first query, at position $p$. After the answer $\Strzero$, the positions $1, \ldots, p$ are known to carry $\Strzero$ and the remaining positions form an arbitrary word of $L_{3}(n - p)$; queries at known positions are useless, so the remaining cost is $f(n - p)$. Symmetrically, after the answer $\Strtwo$, the remaining cost is $f(p - 1)$. After the answer $\Strone$, the state is $\{(a, b) : a \in A, b \in B\}$ with $A = \{0, \ldots, p - 1\}$ and $B = \{p, \ldots, n\}$. In such a product state, a query at a position $q \leq p - 1$ has answer $\Strzero$ or $\Strone$ according to whether $q \leq a$, and a query at a position $q \geq p$ has answer $\Strone$ or $\Strtwo$ according to whether $q \leq b$; thus every query gives information about only one of $a$ and $b$, and it splits the current interval for that coordinate into two parts. Two independent binary searches show that the cost is at most $\lceil \log_{2}\lvert A \rvert \rceil + \lceil \log_{2}\lvert B \rvert \rceil$. Conversely, by induction on the depth $d$ of the tree, every deterministic tree that recognizes the words of a product state with intervals $A$ and $B$ has depth $d \geq \lceil \log_{2}\lvert A \rvert \rceil + \lceil \log_{2}\lvert B \rvert \rceil$. If $d = 0$, the tree performs no query, so $\lvert A \rvert = \lvert B \rvert = 1$. If $d \geq 1$, consider the first query. If it gives no information (one of the two parts is empty), the subtree below it recognizes the same product state and has depth at most $d - 1$, so $d - 1 \geq \lceil \log_{2}\lvert A \rvert \rceil + \lceil \log_{2}\lvert B \rvert \rceil$ by the induction hypothesis. If it concerns $a$ and splits $A$ into two nonempty intervals, one of them, say $A'$, has $\lvert A' \rvert \geq \lvert A \rvert / 2$, so $\lceil \log_{2}\lvert A' \rvert \rceil \geq \lceil \log_{2}\lvert A \rvert \rceil - 1$; the subtree for the product state $(A', B)$ has depth at most $d - 1$, and the induction hypothesis gives $d - 1 \geq \lceil \log_{2}\lvert A \rvert \rceil - 1 + \lceil \log_{2}\lvert B \rvert \rceil$. The case of a query concerning $b$ is symmetric. Hence the cost after the answer $\Strone$ is exactly $\lceil \log_{2} p \rceil + \lceil \log_{2}(n - p + 1) \rceil$, and minimizing over $p$ the maximum over the three answers gives the recursion.

We now prove the closed form. Let $g(n) = \lfloor \log_{2} n \rfloor + \lfloor \log_{2}(n/3) \rfloor + 2$ for $n \geq 2$, and let $e = \lfloor \log_{2} n \rfloor$. If $n < 3 \cdot 2^{e-1}$, then $2^{e-2} < n/3 < 2^{e-1}$ and $g(n) = 2e$; otherwise $2^{e-1} \leq n/3 < 2^{e}$ and $g(n) = 2e + 1$. This gives the two expressions for $g$, and shows that $g$ is nondecreasing. Direct evaluation gives $f(1) = 1$, $f(2) = 2 = g(2)$, and $f(3) = 3 = g(3)$. Let $n \geq 4$, so that $e \geq 2$, and assume that $f(m) = g(m)$ for $2 \leq m < n$. Since $f(0) = 0$ and $f(1) = 1$, this hypothesis yields, for $0 \leq m < n$:
\begin{enumerate}
    \item[(F1)] $f(m) \leq 2e - 1$ if and only if $m < 2^{e}$;
    \item[(F2)] $f(m) \leq 2e - 2$ if and only if $m < 3 \cdot 2^{e-2}$.
\end{enumerate}
For a query position $p$, put $q = n - p + 1$, so that $p + q = n + 1$ and the third term of the recursion is $A(p) = \lceil \log_{2} p \rceil + \lceil \log_{2} q \rceil$.

\emph{Case $2^{e} \leq n < 3 \cdot 2^{e-1}$.} For $p = 2^{e-1}$ we have $q \leq 2^{e}$, so $A(p) \leq 2e - 1$, and $f(p - 1) \leq 2e - 1$ and $f(n - p) \leq 2e - 1$ by (F1), since $p - 1 < 2^{e}$ and $n - p < 2^{e}$. Hence $f(n) \leq 2e$. Conversely, let $p$ be arbitrary and suppose that $f(n - p) \leq 2e - 2$ and $f(p - 1) \leq 2e - 2$. By (F2), $p \leq 3 \cdot 2^{e-2}$ and $q \leq 3 \cdot 2^{e-2}$. Since $p + q \geq 2^{e} + 1$, neither $p$ nor $q$ is at most $2^{e-2}$ (otherwise the other would exceed $3 \cdot 2^{e-2}$), and they are not both at most $2^{e-1}$. Hence $A(p) \geq e + (e - 1) = 2e - 1$. Thus the maximum in the recursion is at least $2e - 1$ for every $p$, and $f(n) \geq 2e$.

\emph{Case $3 \cdot 2^{e-1} \leq n < 2^{e+1}$.} For $p = 2^{e}$ we have $q \leq 2^{e}$, so $A(p) \leq 2e$, and $f(p - 1) \leq 2e - 1$ and $f(n - p) \leq 2e - 1$ by (F1). Hence $f(n) \leq 2e + 1$. Conversely, let $p$ be arbitrary and suppose that $f(n - p) \leq 2e - 1$ and $f(p - 1) \leq 2e - 1$. By (F1), $p \leq 2^{e}$ and $q \leq 2^{e}$. Since $p + q \geq 3 \cdot 2^{e-1} + 1$, neither $p$ nor $q$ is at most $2^{e-1}$, so $A(p) = 2e$. Thus $f(n) \geq 2e + 1$.

This completes the induction. For the final estimate, let $n \geq 2$. If $2^{e} \leq n < 3 \cdot 2^{e-1}$, then $f(n) = 2e \leq 2\log_{2} n < 2e + 2\log_{2}(3/2)$. If $3 \cdot 2^{e-1} \leq n < 2^{e+1}$, then $2\log_{2} n \geq 2e + 2\log_{2}(3/2) > 2e + 1 = f(n)$ and $2\log_{2} n < 2e + 2 < f(n) + 2\log_{2}(3/2)$, using $2\log_{2}(3/2) \approx 1.17 > 1$. In both cases $2\log_{2} n - 2\log_{2}(3/2) < f(n) \leq 2\log_{2} n$.
\end{proof}

\begin{proof}[Proof of Proposition~\ref{prop:exact}]
We use Proposition~\ref{prop:certificates}, the bound of $n$ on all depths, and the following observation: if $w \in \Sigma^{n}$ and, for every position $p$, changing the letter at position $p$ suitably changes whether the word belongs to $L$, then every membership certificate of $w$ contains all $n$ positions, and $\hma_{L}(n) = \hmd_{L}(n) = n$.

$L_{1}$ and $L_{1}'$: both have $\Hom = \infty$, so the recognition depths are $n$ by Theorem~\ref{thm:recognition}(a). The word $\Strzero^{n}$ belongs to both languages, and changing any letter to $\Strtwo$ gives a word outside them, so both membership depths are $n$.

$L_{2} = \Sigma^{*}$: the recognition depths are $n$ by Theorem~\ref{thm:recognition}(a), and the membership depths are $0$ by Theorem~\ref{thm:membership}(b).

$L_{3}$: the recognition depths follow from Propositions~\ref{prop:recursion} and~\ref{prop:sorted} (with $k = 3$, $\min(n, 4)$ covers $n \leq 3$ and $n \geq 4$). For membership with $n \geq 3$, consider $w = \Strzero\Strone^{n-1} \in L_{3}$: changing position $1$ to $\Strtwo$, position $2$ to $\Strtwo$, or a position $p \geq 3$ to $\Strzero$ gives an unsorted word. For $n = 2$, consider $\Strone\Strone$ and change position $1$ to $\Strtwo$ or position $2$ to $\Strzero$.

$L_{4}$: for $n \geq 2$, $L_{4}(n) = \{\Strzero^{n}, \Strzero^{n-1}\Strone, \Strzero^{n-1}\Strtwo, \Strzero^{n-2}\Strone\Strtwo\}$. These four words are distinguished by the letters at positions $n - 1$ and $n$, so $\hrd_{L_{4}}(n) \leq 2$; one query has only three answers, so $\hrd_{L_{4}}(n) \geq 2$; and a certificate of $\Strzero^{n-1}\Strtwo$ must contain position $n$ (to exclude $\Strzero^{n}$) and position $n - 1$ (to exclude $\Strzero^{n-2}\Strone\Strtwo$), so $\hra_{L_{4}}(n) = 2$. For membership with $n \geq 3$, consider $\Strzero^{n-2}\Strone\Strtwo$: changing a letter of the prefix to $\Strtwo$, the letter $\Strone$ to $\Strtwo$, or the letter $\Strtwo$ to $\Strone$ gives a word outside $L_{4}$. For $n = 2$, consider $\Strone\Strtwo$ and change position $1$ to $\Strtwo$ or position $2$ to $\Strone$.

$L_{5}$: $L_{5}(2) = \{\Strzero\Strone, \Strzero\Strtwo, \Strone\Strtwo\}$, $L_{5}(3) = \{\Strzero\Strone\Strtwo\}$, and $L_{5}(n) = \emptyset$ for $n \geq 4$. For $n = 2$, one query does not distinguish the three words (position $1$ does not separate $\Strzero\Strone$ from $\Strzero\Strtwo$, and position $2$ does not separate $\Strzero\Strtwo$ from $\Strone\Strtwo$), and $\Strzero\Strtwo$ needs both positions in any certificate; for membership, changing either letter of $\Strzero\Strone$ to the other letter of the word gives $\Strone\Strone$ or $\Strzero\Strzero$, which are outside $L_{5}$. For $n = 3$, recognition is trivial and every change of $\Strzero\Strone\Strtwo$ leaves $L_{5}$. For $n \geq 4$, all depths are $0$.

Finally, for $n = 1$, $L(1)$ has at least two elements for each of the six languages, so the recognition depths are $1$; the membership depths are $1$ for $L_{1}$ and $L_{1}'$, which do not contain $\Strtwo$, and $0$ for the other languages, which contain all three letters.
\end{proof}

\begin{proof}[Proof of Proposition~\ref{prop:effective}]
Let $L = \Av(F)$. If $F = \emptyset$, then $L = \Sigma^{*}$; otherwise $L^{C} \supseteq F \neq \emptyset$, so $L \neq \Sigma^{*}$. An infinite subword-closed language contains arbitrarily long words, hence words with arbitrarily many occurrences of some letter $\Stra$, and hence $\Stra^{n}$ for every $n$; thus $L$ is infinite if and only if $F \cap \Stra^{*} = \emptyset$ for some $\Stra \in \Sigma$. For letters $\Stra \neq \Strb$, the subwords of $\Stra^{m}\Strb\Stra^{m}$ are the words $\Stra^{x}$ with $x \leq 2m$ and $\Stra^{x}\Strb\Stra^{y}$ with $x, y \leq m$. Hence $\Stra^{m}\Strb\Stra^{m} \in L$ if and only if $m < \mu_{\Stra\Strb}$, where
\[
    \mu_{\Stra\Strb} \;=\; \min\bigl(\{\lceil x/2 \rceil : \Stra^{x} \in F\} \cup \{\max(x, y) : \Stra^{x}\Strb\Stra^{y} \in F\}\bigr), \qquad \min \emptyset = \infty.
\]
Therefore $\Hom(L) = \infty$ if $\mu_{\Stra\Strb} = \infty$ for some pair, and otherwise $\Hom(L) = \max(\{\mu_{\Stra\Strb} - 1 : \Stra \neq \Strb\} \cup \{0\})$. Similarly, the subwords of $\Stra^{m}\Strb^{m}$ are the words $\Stra^{x}\Strb^{y}$ with $x, y \leq m$, so $\Stra^{m}\Strb^{m} \in L$ if and only if $m < \nu_{\Stra\Strb} = \min\{\max(x, y) : \Stra^{x}\Strb^{y} \in F\}$, and $\Het(L)$ is obtained in the same way. For a fixed pair $\Stra \neq \Strb$, deciding whether a word $f \in F$ has one of the forms $\Stra^{x}$, $\Stra^{x}\Strb\Stra^{y}$, or $\Stra^{x}\Strb^{y}$, and reading off $x$ and $y$, takes $O(\lvert f \rvert)$ time, so all values $\mu_{\Stra\Strb}$ and $\nu_{\Stra\Strb}$ are computed in $O(k^{2}\lVert F \rVert)$ time, and the finiteness test takes $O(k\lVert F \rVert)$ time. The minimality of $F$ is not used in this argument.
\end{proof}

\section{Comparison with the binary case}\label{sec:comparison}

The main tool of \cite{Moshkov2023} is \cite[Lemma~1]{Moshkov2023}: if $\Hom(L) = m < \infty$, every word of a binary subword-closed language has the form $w_{1}\Stra^{i}w_{2}\bar{\Stra}^{j}w_{3}$ with $\lvert w_{1} \rvert, \lvert w_{2} \rvert, \lvert w_{3} \rvert \leq 2m$. Its proof is a case analysis according to whether each letter has few occurrences to its left or to its right, which uses that there are only two letters. Building on it, the logarithmic upper bound in \cite[Theorem~1(b)]{Moshkov2023} splits a word into $X_{1}X_{2}X_{3}AY_{3}Y_{2}Y_{1}$ (written $L_{1}L_{2}L_{3}AR_{3}R_{2}R_{1}$ in \cite{Moshkov2023}) with factors $X_{i}, Y_{i}$ of length $t = 2m$, distinguishes four cases according to whether $X_{2}$ and $Y_{2}$ are constant, and, in the main case, locates the \emph{single} region where the word changes from $\Stra$ to $\bar{\Stra}$ by binary search over blocks of length $t$. For $k \geq 3$, neither step works: there may be up to $k$ long blocks (Remark~\ref{rem:lemma1fails}), and hence up to $k - 1$ regions of change that must be located simultaneously. Our argument differs in two respects. First, Lemma~\ref{lem:structure} is proved by locating, for each heavy letter separately, the factor between its $(m + 1)$-th occurrence from the left and from the right, which avoids the case analysis. Second, the procedure of Lemma~\ref{lem:logdepth} does not need to know the number or the order of the blocks: it refines all regions of change at once, and fact~1 lets it skip every interval whose endpoints are constant with the same letter. For $k = 2$ this gives another proof of \cite[Theorem~1]{Moshkov2023}.

The quantitative comparison for $k = 2$ is as follows. In the bounded case~(c), \cite{Moshkov2023} obtains $\hrd_{L}(n) \leq p^{2}$ with $p = 2^{3t'+3}(2t' + 4)$ and $t' = 2\max(\Hom(L), \Het(L))$, which is exponential in $\Hom(L)$ and $\Het(L)$, whereas Lemma~\ref{lem:bounded-depth} gives $\hrd_{L}(n) \leq 2T + 2 = 8\Hom(L) + 2\Het(L) + 6$, which is linear. So already for $k = 2$ the bound is improved by an exponential factor. The bound $\lvert L(n) \rvert \leq (T + 1)2^{T+2}$ of Corollary~\ref{cor:bounded} does not exceed Moshkov's bound $p$ whenever $\max(\Hom(L), \Het(L)) \geq 1$ (with equality for $\Hom(L) = \Het(L) = 1$); for $\Hom(L) = \Het(L) = 0$ it is larger ($48$ versus $32$).

In case~(b), the proof of \cite[Theorem~1(b)]{Moshkov2023} represents every word of $L$ as $w_{1}\Stra^{i}w_{2}\bar{\Stra}^{j}w_{3}$ where ``the length of $w_{k}$ is at most $t = 2\Hom(L)$'', considers $n \geq 10t$, and states that ``in case (iv), we make at most $t\log_{2}(n/t) + 7t$ queries'' and that ``as $c$, we can use the number $7t$'' for the nondeterministic depth. These expressions degenerate when $\Hom(L) = 0$, i.e., $t = 0$ (for instance, $L = \{\Strzero^{i}\Strone^{j}\}$ has $\Hom(L) = 0$, $\hra_{L}(n) = 2$, and $\hrd_{L}(n) = \lceil \log_{2}(n + 1) \rceil$), so the argument of \cite{Moshkov2023} has to be read with $t \geq 1$ in that case, e.g., $t = \max(1, 2\Hom(L))$. For $m = \Hom(L) \geq 1$, the comparison is as follows. For nondeterministic recognition, \cite{Moshkov2023} obtains $7t = 14m$, and Lemma~\ref{lem:certificate} gives $8m + 6$, which is not larger. For deterministic recognition, \cite{Moshkov2023} obtains $2m\log_{2}(n/2m) + 14m$, whereas Lemma~\ref{lem:logdepth} gives $(2m + 1)(2 + 20\lceil \log_{2} n \rceil)$. As $n \to \infty$, the ratio of the two bounds tends to $20(2m + 1)/(2m)$, which equals $30$ for $m = 1$ and tends to $20$ as $m \to \infty$. In this case our bound is worse by a constant factor. For $m = 0$, Proposition~\ref{prop:hom0} gives $2 + \lceil \log_{2} n \rceil$, which exceeds the optimal value for $\{\Strzero^{i}\Strone^{j}\}$ by at most $2$. Table~\ref{tab:comparison} lists the bounds of \cite{Moshkov2023} together with ours, for $k = 2$ and for general $k$.

\begin{table}[ht]
\centering
\caption{Bounds of \cite{Moshkov2023} for $k = 2$ and bounds of this paper. Here $m = \Hom(L)$, $h = \Het(L)$, $T = k(2m + 1) + (k - 1)h$, and $p = 2^{3t'+3}(2t' + 4)$ with $t' = 2\max(m, h)$. The bounds of \cite{Moshkov2023} in case~(b) require $m \geq 1$.}
\label{tab:comparison}
\small
\begin{tblr}{colspec = {l | c | c c}, row{1} = {font=\bfseries}, cell{1}{2-4} = {c}, hline{1,Z} = {0.08em}, hline{2} = {0.05em}, rowsep = 2pt}
Quantity & {\cite{Moshkov2023}, $k = 2$} & {This paper, $k = 2$} & {This paper, $k \geq 2$} \\
Total gap length & $6m$ & $4m + 2$ & $k(2m + 1)$ \\
$\hra_{L}(n)$, case (b) & $14m$ & $8m + 6$ & $k(4m + 3)$ \\
$\hrd_{L}(n)$, case (b) & $2m\log_{2}\frac{n}{2m} + 14m$ & $(2m + 1)(2 + 20\lceil \log_{2} n \rceil)$ & $(2m + 1)(2 + 10k\lceil \log_{2} n \rceil)$ \\
$\hrd_{L}(n)$, $\Hom(L) = 0$ & not defined & $2 + \lceil \log_{2} n \rceil$ & $2 + (k - 1)\lceil \log_{2} n \rceil$ \\
$\hrd_{L}(n)$, case (c) & $p^{2}$ & $8m + 2h + 6$ & $2T + 2$ \\
$\lvert L(n) \rvert$, case (c) & $p$ & $(T + 1)2^{T+2}$ & $(T + 1)k^{T+2}$ \\
\end{tblr}
\end{table}

\section{Conclusions and open problems}\label{sec:conclusions}

Moshkov's classification of binary subword-closed languages holds unchanged for every alphabet with $k \geq 2$ letters, with the same five complexity classes. The proofs use only properties of subwords, and no automaton for $L$ is needed. The alphabet size affects the constants: by Proposition~\ref{prop:sorted}, the constant of the logarithmic term grows at least like $(k - 1)/\log_{2} k$.

The following problems remain open.
\begin{enumerate}
    \item In case~(b), is the multiplicative dependence on $\Hom(L)$ in Lemma~\ref{lem:logdepth} necessary? For $L = \mathord{\downarrow}\{\Strzero^{m}\Strone\Strzero^{m}\} \cup \{\Strzero^{i}\Strone^{j}\}$ we have $\Hom(L) = m$ and $\Het(L) = \infty$, but $L(n) = \{\Strzero^{i}\Strone^{n-i}\}$ for $n > 2m + 1$, so $\hrd_{L}(n) = \lceil \log_{2}(n + 1) \rceil$ for such $n$. We conjecture that $\hrd_{L}(n) \leq c(k)\log n + d(k, \Hom(L))$ for all $L$ in class $\mathcal{C}_{3}$. For $\Hom(L) = 0$ this holds with $c(k) = k - 1$ (in base $2$) by Proposition~\ref{prop:hom0}.
    \item Does $\lim_{n \to \infty} \hrd_{L}(n)/\log_{2} n$ exist for every $L$ in class $\mathcal{C}_{3}$? For $L_{3}$ it exists and equals $2$ by Proposition~\ref{prop:recursion}. For $\Lsort$ over $k$ letters we conjecture that it equals $k - 1$.
    \item In case~(c), is $\hrd_{L}(n)$ eventually constant? Lemma~\ref{lem:shape} suggests that, for large $n$, the set of possible pairs $(u, v)$ in the representation $u\Strc^{i}v$ stabilizes.
    \item Do $\hmd_{L}(n)$ and $\hma_{L}(n)$ coincide for every subword-closed language $L$ and every $n$? They coincide for all examples considered here.
    \item What is the complexity of determining the class of $L$ from a deterministic finite automaton accepting $L$? From a finite set of forbidden subwords it is polynomial (Proposition~\ref{prop:effective}).
\end{enumerate}
A further direction is to extend the approach to other classes of formal languages, such as the factorial languages considered in \cite{Moshkov2022} and \cite[pp.~223--237]{DurdymyradovMoshkovOstonov2024}.

\section*{Acknowledgments}

The author thanks Mikhail Moshkov for introducing the author to decision trees for formal languages.

\section*{Data availability}

The Python scripts used for the computations in Section~\ref{sec:computation} (exact minimum depths, evaluation of the recursion of Proposition~\ref{prop:recursion}, and randomized checks of the lemmas), together with an explanation that lists the exact commands regenerating every number in Tables~\ref{tab:exact} and~\ref{tab:recursion}, will be provided as supplementary material and will be deposited in a public repository. Until then, they are available from the author on request.

\section*{Use of generative AI}

During the preparation of this work the author used Claude (Anthropic) to draft and revise parts of the text, to write the Python scripts used for the computational verification, and to prepare the LaTeX source and bibliography. The author reviewed, verified, and edited all content and takes full responsibility for it.

\end{document}